%% file: main.tex
\documentclass[10pt,journal,compsoc]{IEEEtran}
\ifCLASSOPTIONcompsoc
\else
\fi
\ifCLASSINFOpdf
\else
\fi

\usepackage{amssymb}
\usepackage{array}

\usepackage{epsfig,endnotes,comment}

\usepackage[hyphens]{url}
\usepackage{cite}

\usepackage{multirow}
\usepackage{flushend}

\newcommand{\cm}{\checkmark}
\usepackage[usenames,dvipsnames]{xcolor}

\newcommand{\edit}[1]{{\color{Black}#1}}

\newcommand{\sss}[1]{~\\ \noindent \textsf{#1:}}

\newcommand{\secTestbeds}{Network Testbeds}

\begin{document}
%
% paper title
% can use linebreaks \\ within to get better formatting as desired
% Do not put math or special symbols in the title.
\title{Survey of End-to-End Mobile Network Measurement Testbeds, Tools, and Services}
%
%
% author names and IEEE memberships
% note positions of commas and nonbreaking spaces ( ~ ) LaTeX will not break
% a structure at a ~ so this keeps an author's name from being broken across
% two lines.
% use \thanks{} to gain access to the first footnote area
% a separate \thanks must be used for each paragraph as LaTeX2e's \thanks
% was not built to handle multiple paragraphs
%
%
%\IEEEcompsocitemizethanks is a special \thanks that produces the bulleted
% lists the Computer Society journals use for "first footnote" author
% affiliations. Use \IEEEcompsocthanksitem which works much like \item
% for each affiliation group. When not in compsoc mode,
% \IEEEcompsocitemizethanks becomes like \thanks and
% \IEEEcompsocthanksitem becomes a line break with idention. This
% facilitates dual compilation, although admittedly the differences in the
% desired content of \author between the different types of papers makes a
% one-size-fits-all approach a daunting prospect. For instance, compsoc 
% journal papers have the author affiliations above the "Manuscript
% received ..."  text while in non-compsoc journals this is reversed. Sigh.

\author{Utkarsh~Goel, %~\IEEEmembership{Student Member,~IEEE,}
        Mike~P.~Wittie, %~\IEEEmembership{Member,~IEEE,}
        Kimberly C. Claffy, %~\IEEEmembership{Member,~IEEE}\\
        and~Andrew Le
\IEEEcompsocitemizethanks{\IEEEcompsocthanksitem U.~Goel and M.P.~Wittie are with the Computer Science Department, Montana State University, Bozeman, MT 59717.\protect\\
% note need leading \protect in front of \\ to get a newline within \thanks as
% \\ is fragile and will error, could use \hfil\break instead.
E-mail: \texttt{utkarsh.goel, mwittie@cs.montana.edu}}% <-this % stops an unwanted space
\IEEEcompsocitemizethanks{\IEEEcompsocthanksitem KC Claffy is with UCSD/CAIDA, La~Jolla, CA 92093\protect\\
% note need leading \protect in front of \\ to get a newline within \thanks as
% \\ is fragile and will error, could use \hfil\break instead.
E-mail: \texttt{kc@caida.com}}% <-this % stops an unwanted space
\IEEEcompsocitemizethanks{\IEEEcompsocthanksitem A.Le is with Mintybit, Santa Barbara, CA 93111\protect\\
% note need leading \protect in front of \\ to get a newline within \thanks as
% \\ is fragile and will error, could use \hfil\break instead.
E-mail: \texttt{andrew@mintybit.com}}% <-this % stops an unwanted space
%\thanks{Manuscript received April 19, 2005; revised December 27, 2012.}
}

\IEEEtitleabstractindextext{%
\begin{abstract}
\input{abstract}
\end{abstract}

% Note that keywords are not normally used for peerreview papers.
}

% make the title area
\maketitle

% To allow for easy dual compilation without having to reenter the
% abstract/keywords data, the \IEEEtitleabstractindextext text will
% not be used in maketitle, but will appear (i.e., to be "transported")
% here as \IEEEdisplaynontitleabstractindextext when the compsoc 
% or transmag modes are not selected <OR> if conference mode is selected 
% - because all conference papers position the abstract like regular
% papers do.
\IEEEdisplaynontitleabstractindextext
% \IEEEdisplaynontitleabstractindextext has no effect when using
% compsoc or transmag under a non-conference mode.

% For peer review papers, you can put extra information on the cover
% page as needed:
% \ifCLASSOPTIONpeerreview
% \begin{center} \bfseries EDICS Category: 3-BBND \end{center}
% \fi
%
% For peerreview papers, this IEEEtran command inserts a page break and
% creates the second title. It will be ignored for other modes.
\IEEEpeerreviewmaketitle

%\section{Introduction}
% Computer Society journal papers do something a tad strange with the very
% first section heading (almost always called "Introduction"). They place it
% ABOVE the main text! IEEEtran.cls currently does not do this for you.
% However, You can achieve this effect by making LaTeX jump through some
% hoops via something like:
%
%\ifCLASSOPTIONcompsoc
%  \noindent\raisebox{2\baselineskip}[0pt][0pt]%
%  {\parbox{\columnwidth}{\section{Introduction}\label{sec:introduction}%
%  \global\everypar=\everypar}}%
%  \vspace{-1\baselineskip}\vspace{-\parskip}\par
%\else
%  \section{Introduction}\label{sec:introduction}\par
%\fi
%
% Admittedly, this is a hack and may well be fragile, but seems to do the
% trick for me. Note the need to keep any \label that may be used right
% after \section in the above as the hack puts \section within a raised box.

\input{introduction}

\input{goals}

\input{testbeds}

\input{tools}

\input{services}

\input{conclusions}
\input{acknowledgements}

% Can use something like this to put references on a page
% by themselves when using endfloat and the captionsoff option.
\ifCLASSOPTIONcaptionsoff
  \newpage
\fi

% trigger a \newpage just before the given reference
% number - used to balance the columns on the last page
% adjust value as needed - may need to be readjusted if
% the document is modified later
%\IEEEtriggeratref{8}
% The "triggered" command can be changed if desired:
%\IEEEtriggercmd{\enlargethispage{-5in}}

% references section

% can use a bibliography generated by BibTeX as a .bbl file
% BibTeX documentation can be easily obtained at:
% http://www.ctan.org/tex-archive/biblio/bibtex/contrib/doc/
% The IEEEtran BibTeX style support page is at:
% http://www.michaelshell.org/tex/ieeetran/bibtex/
%\bibliographystyle{IEEEtran}
% argument is your BibTeX string definitions and bibliography database(s)
%\bibliography{IEEEabrv,../bib/paper}
%
% <OR> manually copy in the resultant .bbl file
% set second argument of \begin to the number of references
% (used to reserve space for the reference number labels box)
%\begin{thebibliography}{1}
%
%\bibitem{IEEEhowto:kopka}
%H.~Kopka and P.~W. Daly, \emph{A Guide to \LaTeX}, 3rd~ed.\hskip 1em plus
%  0.5em minus 0.4em\relax Harlow, England: Addison-Wesley, 1999.
%
%\end{thebibliography}

\footnotesize 
\bibliographystyle{IEEEtran}
\bibliography{msr_tools_survey}
%\bibliography{../references/msr_tools_survey}

% biography section
% 
% If you have an EPS/PDF photo (graphicx package needed) extra braces are
% needed around the contents of the optional argument to biography to prevent
% the LaTeX parser from getting confused when it sees the complicated
% \includegraphics command within an optional argument. (You could create
% your own custom macro containing the \includegraphics command to make things
% simpler here.)
%\begin{IEEEbiography}[{\includegraphics[width=1in,height=1.25in,clip,keepaspectratio]{mshell}}]{Michael Shell}
% or if you just want to reserve a space for a photo:

%\begin{IEEEbiography}{Utkarsh Goel}
%Biography text here.
%\end{IEEEbiography}

% if you will not have a photo at all:
%\begin{IEEEbiographynophoto}{Mike Wittie}
%Biography text here.
%\end{IEEEbiographynophoto}

% insert where needed to balance the two columns on the last page with
% biographies
%\newpage

%\begin{IEEEbiographynophoto}{Qing Yang}
%Biography text here.
%\end{IEEEbiographynophoto}

% You can push biographies down or up by placing
% a \vfill before or after them. The appropriate
% use of \vfill depends on what kind of text is
% on the last page and whether or not the columns
% are being equalized.

%\vfill

% Can be used to pull up biographies so that the bottom of the last one
% is flush with the other column.
%\enlargethispage{-5in}

% that's all folks
\end{document}

%% file: abstract.tex
Mobile (cellular) networks enable innovation, but can also stifle it and lead to user frustration when network performance falls below expectations.
As mobile networks become the predominant method of Internet access, developer, research, network operator, and regulatory communities have taken an increased interest in measuring end-to-end mobile network performance to, among other goals, minimize negative impact on application responsiveness.
In this survey we examine current approaches to end-to-end mobile network performance measurement, diagnosis, and application prototyping.
We compare available tools and their shortcomings with respect to the needs of researchers, developers, regulators, and the public.
We intend for this survey to provide a comprehensive view of currently active efforts and some auspicious directions for future work in mobile network measurement and mobile application performance evaluation.

%% file: introduction.tex
\section{Introduction} 

\IEEEPARstart{M}{obile} (cellular) network applications deliver interactive services, generally supported by back-end logic deployed on cloud infrastructure.
These applications support a wide breadth of functionality, such as live video, social gaming, communication services, and augmented reality~\cite{14USTREAM, Codorniou12What, Rogers14What, Marupaka14Future}.
Future services will increasingly leverage cloud-based datasets and processing power for innovative applications of live speech translation, real-time video analysis, or other computationally intensive tasks~\cite{Shubber13Microsoft, Rimington13Leave}.
As the frequency of interactions between mobile devices and back-end servers increases, application responsiveness will be increasingly tightly coupled with end-to-end network performance.

To innovate in the interactive mobile application space, developers deploy communication protocols with sophisticated data delivery techniques that support responsive communications under a range of network conditions~\cite{Grigorik13High, Zhang11Pangolin:, Agarwal09Matchmaking, Erman13Towards, Butler08Proposal}.
These techniques are not always sufficient and developers are sometimes forced to redesign application functionality to mask poor network performance.
However, these latter optimizations require detailed network performance data that is often not readily available, which results in challenges across the cellular ecosystem. 
For example, \textbf{developers} face the undesirable choice of evaluating performance of their mobile applications in limited private deployments that lack geographic diversity, or distributing their code to users without adequate testing~\cite{mobitest:webinterface, 14Perfecto}.
\textbf{Researchers} lack network performance data, or tools to acquire such data, in order to rapidly test hypotheses and focus on realistic network performance problems.
\edit{\textbf{Network operators} need to monitor and troubleshoot end-to-end network performance without degrading base station throughput.}
Finally, \textbf{regulators} have a limited view of network performance, especially with respect to traffic shaping by network providers, impeding their ability to tackle performance challenges and roadblocks for sustained innovation in the mobile space~\cite{Johnston13Measuring, contentprovider:loser}. 

This paper provides a comparative analysis of currently available network measurement platforms for end-to-end mobile network measurement, monitoring, and experimentation.
\edit{We further categorize measurement platforms as 
research testbeds for network experimentation, 
extensible distributed measurement tools, 
and services for widespread monitoring of networks performance.
In the following sections describe the most salient features of each platform, and how some features differ across them.
Table~\ref{tbl:exp-tools} compares the testbeds and tools in terms of their experimentation flexibility, device selection criteria, resource protection, and other features.
}

\input{table.tex}
\input{legend.tex}

%@@ %@@ can you include estimated deployment numbers in the table, as another row?
%@@ mwittie 5/5/15: the difficulty is that deployment numbers we have are from different dates and so hard to compare. They would also take up a lot of space in the table. We discuss the numbers we have in the text.

Based on our review of current measurement efforts, we observe that although existing approaches comprise only a patchwork of needed functionality, they already generate powerful insights to guide development, research, and regulatory actions. 
However, in spite of the relative maturity of several measurement platforms, daunting problems remain including support for wide-scale application prototyping and deployment, detection of traffic shaping, and long-term network performance monitoring.
Most existing mobile measurement tools have been developed in isolation, and one motivation for this survey is to foster more concerted and cooperative efforts at standardization of measurement libraries, privacy policies, and technology exchange~\cite{Zevenberger13Ethical, mobiperf:details, 14AIMS, 14What}.

The rest of this paper is organized as follows.
Section~\ref{sec:goals} reviews goals of end-to-end mobile network measurement.
Sections~\ref{sec:testbeds}, \ref{sec:tools}, and~\ref{sec:services} respectively discuss measurement testbeds, tools, and services for end-to-end mobile network measurement.
Section~\ref{sec:conclusions} presents directions for future work and concluding thoughts.

%% file: table.tex
\begin{table*}
\centering
\begin{tabular}{|>{\arraybackslash}p{3.5cm}|>{\centering\arraybackslash}p{0.2cm}|>{\centering\arraybackslash}p{0.3cm}|>{\centering\arraybackslash}p{0.2cm}|>{\centering\arraybackslash}p{0.3cm}|>{\centering\arraybackslash}p{0.2cm}|>{\centering\arraybackslash}p{0.2cm}|>{\centering\arraybackslash}p{0.3cm}|>{\centering\arraybackslash}p{0.2cm}|>{\centering\arraybackslash}p{0.2cm}|>{\centering\arraybackslash}p{0.35cm}|>{\centering\arraybackslash}p{0.2cm}|>{\centering\arraybackslash}p{0.2cm}|>{\centering\arraybackslash}p{0.45cm}|>{\centering\arraybackslash}p{0.45cm}|>{\centering\arraybackslash}p{0.45cm}|>{\centering\arraybackslash}p{0.3cm}|>{\centering\arraybackslash}p{0.2cm}|>{\centering\arraybackslash}p{0.2cm}|>{\centering\arraybackslash}p{0.2cm}|>{\centering\arraybackslash}p{0.2cm}|}
\hline
\multirow{3}{*}{~}                               & \multicolumn{6}{c|}{\textbf{Network Testbeds}}                                                                                                                                                                                                                               & \multicolumn{7}{c|}{\textbf{Network Tools}}                                                                                                                                                                                                                                          & \multicolumn{7}{c|}{\textbf{Network Services}}                                                                                                                                                                                                                                                                                                           \\ \cline{2-21} 
                                                 & \multicolumn{3}{c|}{\textbf{Uncurated}}                                                                                          & \multicolumn{3}{c|}{\textbf{Curated}}                                                                                             & \multicolumn{5}{c|}{\textbf{Standalones}}                                                                                                                                                & \multicolumn{2}{c|}{\textbf{Libraries}}                                            & \multicolumn{4}{c|}{\textbf{\begin{tabular}[c]{@{}c@{}} Network \\ Monitoring\end{tabular}}}                                                                                                       & \multicolumn{3}{c|}{\textbf{\begin{tabular}[c]{@{}c@{}}Network \\ Discovery \& \\ Diagnosis\end{tabular}}}                  \\ \cline{2-21} 
& \rotatebox[origin=l]{90}{\textbf{MITATE}}
& \rotatebox[origin=l]{90}{\textbf{Seattle}}
& \rotatebox[origin=l]{90}{\textbf{PhantomNet}}
& \rotatebox[origin=l]{90}{\textbf{PhoneLab}}
& \rotatebox[origin=l]{90}{\textbf{SciWiNet}}
& \rotatebox[origin=l]{90}{\textbf{LiveLabs}}
& \rotatebox[origin=l]{90}{\textbf{FCC SpeedTest}}
& \rotatebox[origin=l]{90}{\textbf{WindRider}}
& \rotatebox[origin=l]{90}{\textbf{MySpeedTest}}
& \rotatebox[origin=l]{90}{\textbf{Mobitest}}
& \rotatebox[origin=l]{90}{\textbf{RILAnalyzer}}
& \rotatebox[origin=l]{90}{\textbf{Mobiperf}}
& \rotatebox[origin=l]{90}{\textbf{ALICE}}
& \rotatebox[origin=l]{90}{\textbf{Ookla SpeedTest~}}
& \rotatebox[origin=l]{90}{\textbf{RadioOpt}}
& \rotatebox[origin=l]{90}{\textbf{OpenSignal}}
& \rotatebox[origin=l]{90}{\textbf{NetPerform}}
& \rotatebox[origin=l]{90}{\textbf{NDT}}
& \rotatebox[origin=l]{90}{\textbf{Netalyzr}}
& \rotatebox[origin=l]{90}{\textbf{PortoLan}} \\ \hline
\multicolumn{21}{|l|}{\textbf{Measurement Capabilities}}                                                                                                                                                                                                                                                                                                                                                              \\ \hline
Traffic shaping/DPI       & \cm        & \cm        &               & \cm         &            &            &                & \cm       &             &          &  &                      &                  &                           &                     &                     &                     &                               &                                    & \cm \\ \hline
Active measurements                             &  \cm          &  \cm          & \cm              & \cm         &   \cm         & \cm        &   \cm             & \cm       & \cm         &    \cm      &  & \cm                  & \cm              & \cm                          & \cm                    & \cm                 & \cm                 & \cm                              &   \cm                                 & \cm                                               
\\ \hline
Passive measurements                             &            &            &               & \cm         &            & \cm        &                & \cm       & \cm         &          & \cm & \cm                  & \cm              &                           &                     & \cm                 & \cm                 &                               &                                    &                                    \\ \hline
Measurement data publicly available              & 2          &            &               & \cm         &            &            & 2              &           &             &          & & \cm                  & \cm              &                           &                     & \cm                 &                     & \cm                           &                                    &                                    \\ \hline
Custom packet content                            & \cm        & \cm        &               & \cm         &            &            &                &           &             & \cm    &  &                      &                  & 1                         &                     &                     &                     &                               &                                    &                                    \\ \hline
Peer-to-peer traffic                             & 2          & \cm        & \cm           &             &            &            &                &           &             &          & &                      &                  &                           &                     &                     &                     &                               &                                    &                                    \\ \hline
ICMP traceroutes                                 & 2          &            &               & \cm         &            &            &                &           &             &          & & \cm                  & \cm              &                           &                     &                     &                     &                               & 3                                  & \cm                                \\ \hline
Programmable execution environment               & 4          & \cm        & \cm           & \cm         &            & \cm        &                &           &             &          & &                      & \cm              &                           &                     &                     &                     &                               & \cm                                &                                    \\ \hline
Access to mobile device sensor data              & \cm        & 2          &               & \cm         & \cm        & \cm        &                & 5         & 6           &         & 7 & 5,6                  & 5,7              & 5                         & 5,6, 7              & 5,6, 7              & 5,6, 7              &                               & 5,7                                & \cm                                \\ \hline
Experiments can be scheduled on specific clients & \cm        & \cm        & \cm           & \cm         &            & \cm        &                &           &             & \cm     &  &                      & \cm              &                           &                     &                     &                     &                               & \cm                                &                                    \\ \hline
IPv6 support                                     & 2          & 2          &               &             &            &            &                &           &             &          & & \cm                  &                  &                           &                     &                     &                     &                               & \cm                                & \cm                                \\ \hline
Allow traffic on ports $<$ 1024                        & 3          & 3          &               & \cm         &            &            &                &           &             &          & &                      &                  &                           &                     &                     &                     &                               &                                    &                                    \\ \hline
Reports network problems                         &            &            &               &             &            &            &                &           &             &         & & \cm                  &                  &                           &                     &                     &                     & \cm                           & \cm                                &                                    \\ \hline
Supported mobile OS platform                     & 8          & 8,9, 12    & 8             & 8           & 8          & 8,9        & 8,9            & 10        & 8           & 8,9, 11  & 8 & 8                    & 8                & 8,9, 10,13                & 8,9, 10,11          & 8,9                 & 8,9                 & 8                             & 8                                  & 8                                  \\ \hline
Network coverage map                             &            &            &               &             &            &            &                &           &             &          & &                      &                  &                           &                     & \cm                 & \cm                 &                               &                                    &                                    \\ \hline
\multicolumn{21}{|l|}{\textbf{Device Selection Criteria}}                                                                                                                                                                                                                                                                                                                                                                                  \\ \hline
Geographic location                              & \cm        & \cm        &               & \cm         & \cm        & \cm        &                &           &             &         & &                      &                  &                           &                     & Q                   &                     &                               &                                    &                                    \\ \hline
Device model                                     & \cm        &            &               &             &            &            &                &           &             & \cm    &  & \cm                  &                  &                           &                     &                     &                     &                               &                                    &                                    \\ \hline
Device type (GSM/CDMA)                           &            &            &               &             &            &            &                &           &             &      &    &                      &                  &                           &                     & Q                   &                     &                               &                                    &                                    \\ \hline
Battery charge level                             & \cm        & \cm        &               & \cm         &            &            &                &           &             &      &    &                      &                  &                           &                     & Q                   &                     &                               &                                    &                                    \\ \hline
Carrier signal strength                          & \cm        &            &               & \cm         &            &            &                &           &             &      &    &                      &                  &                           &                     & Q                   &                     &                               &                                    &                                    \\ \hline
Network carrier                                  & \cm        &            &               &             &            &            &                &           &             &       &   &                      & \cm              &                           &                     & Q                   &                     &                               &                                    &                                    \\ \hline
Network type (Wi-Fi/Cellular)                    & \cm        &            &               & \cm         &            &            &                &           &             &        &  &                      & \cm              &                           &                     & Q                   &                     &                               &                                    &                                    \\ \hline
Time of day                                      & \cm        & \cm        & \cm           & \cm         &            & \cm        &                &           &             & \cm    &  &                      & \cm              &                           &                     &                     &                     &                               &                                    &                                    \\ \hline
\multicolumn{21}{|l|}{\textbf{Resource Usage Limit}}                                                                                                                                                                                                                                                                                                                                                                                       \\ \hline
Transmission rate                                &            & \cm        &               &             &            &            &                &           &             &        &  &                      &                  &                           &                     &                     &                     &                               &                                    &                                    \\ \hline
Bandwidth cap                                    & \cm        & \cm        &               &             & \cm        &            & \cm            &           & \cm         &         &  & \cm                  & \cm              &                           & \cm                 &                     &                     &                               &                                    & \cm                                \\ \hline
Minimum battery charge                           & \cm        & 2          &               &             &            &            &                &           &             &        &   & \cm                  &                  &                           &                     &                     &                     &                               &                                    & \cm                                \\ \hline
Port restrictions                                &            & \cm        &               &             & \cm        &            &                &           &             &       &   &                      &                  &                           &                     &                     &                     &                               &                                    &                                    \\ \hline
\multicolumn{21}{|l|}{\textbf{Misc.}}                                                                                                                                                                                                                                                                                                                                                                                                      \\ \hline
Measurement scheduling API                       & \cm        & \cm        & \cm           &             &            & \cm        &                &           &             &       &   & \cm                  & \cm              &                           &                     &                     &                     &                               &                                    &                                    \\ \hline
Supports devices behind NAT/Wi-Fi                & \cm        & \cm        &               & \cm         & \cm        &            & \cm            &           & \cm         & \cm    &  & \cm                  & \cm              & \cm                       & \cm                 & \cm                 & \cm                 & \cm                           & \cm                                & \cm                                \\ \hline
Requires rooted phones                           &            &            & \cm           & \cm         &            &            &                &           &             & \cm   & \cm   &                      &                  &                           &                     &                     &                     &                               &                                    &                                    \\ \hline
Open to public                                   & \cm        & \cm        & \cm           & \cm         & \cm        &            & \cm            & \cm       & \cm         & \cm    & \cm  & \cm                  & \cm              & \cm                       & \cm                 & \cm                 & \cm                 & \cm                           & \cm                                & \cm                 \\ \hline
User incentive model                             & R          & R,A        &               & U           &            & S,U        & C              & C         & C           & A    &    & C,S                  & S                & C                         & C,S                 & C,S                 & C,S                 & S                             & S                                  & A                                  \\ \hline
Experiments require IRB approval                 &            &            &               & \cm         &            & \cm        &                &           &             &       &   &                      &                  &                           &                     &                     &                     &                               &                                    &                                    \\ \hline
Open-source                                      & \cm        & \cm        &               &             &            &            & \cm            & \cm       & \cm         & \cm   & \cm   & \cm                  &                  &                           &                     &                     &                     & \cm                           &                                    &                                    \\ \hline
Currently active                                 & D          & \cm        & D             & \cm         & D          & D          & \cm            & \cm       & \cm         & \cm    & \cm  & \cm                  & D                & \cm                       & \cm                 & \cm                 & \cm                 & \cm                           &                                    & \cm                                \\ \hline
Records hardware specs                           & O          &            &               &             & \cm        &            & \cm            & \cm       &             & \cm   &   & \cm                  &                  &                           & \cm                 & \cm                 & \cm                 &                               &                                    &                                    \\ \hline
Records hardware performance                     & O          & O          &               & O           &            &            &                &           & \cm         & 6      &  &                      &                  &                           & \cm                 & \cm                 &                     &                               &                                    &                                    \\ \hline
\end{tabular}
\end{table*}

%% file: legend.tex
\begin{table*}
\begin{tabular}{l}
\textbf{Legends:}                                                 \\~\\
1 -- measurements can be directed to specific Web servers.       \\
2 -- planned functionality.                                      \\
3 -- on rooted phones only.                                      \\
4 -- through multiple experiment rounds on the same device.      \\
5 -- GPS readings.                                               \\
6 -- battery readings.                                           \\
7 -- radio state.                                                \\
8 -- Android.                                                    \\
9 -- iOS.                                                        \\
10 -- Windows.                                                   \\
11 -- Blackberry.                                                \\
12 -- Nokia.                                                     \\
13 -- Amazon FireOS                                              \\
A -- user Altruism to support measurement capacity.              \\
D -- under Deployment.                                           \\
O -- Optionally.                                                 \\
C -- user Curiosity to understand their own network performance. \\
Q -- only at query time.                                         \\
R -- Reciprocal (tit-for-tat).                                   \\
S -- provides Service to clients other than measurement data.    \\
                                                                
\end{tabular}
\\~\\
\caption{Experimentation flexibility matrix of end-to-end measurement testbeds, tools, and services.}
\label{tbl:exp-tools}
\end{table*}

%% file: goals.tex
\section{Goals of end-to-end mobile\\ network measurement}
\label{sec:goals}

The 2014 CAIDA workshop on Active Internet Measurements~(AIMS 2014) brought together developers, researchers, network operators, and regulators interested in mobile (and wireless) network performance~\cite{14AIMS}.
Although these communities share the goals of understanding and improving performance of current mobile networks, they focus on different metrics, and thus the tools they produce (Section~\ref{sec:testbeds}) take different approaches.
\looseness-1

\subsection{Developers' View of Network Performance}

Developers want to provide a responsive application experience to their users.
Although much of the delay experienced by user requests is due to back-end processing and front-end rendering~\cite{Erman13Towards}, as hardware and software processing speed improves, network latency becomes a dominant concern.
Moreover, network latency does not necessarily benefit from advances in communication technology.
Internet Service Providers~(ISPs) typically engineer their networks to minimize forwarding costs, which is not always aligned with minimal end-to-end latency.
Specifically, ISPs may direct traffic onto inexpensive but circuitous routes, which inflates hop counts and path latency~\cite{goel:mitatepaper}. 
ISPs may also configure cellular schedulers to delay transmissions until carrier channel conditions are favorable~\cite{Winstein13Stochastic}.
Additionally, mobile networks remain bandwidth-constrained, which motivates ISPs to deploy traffic shaping mechanisms on video streaming and P2P traffic to increase the usable bandwidth for all mobile users~\cite{GigacomTrafficshaping11}.
Traffic shaping can induce high latency that impedes the performance of dynamic content applications such as interactive Web, live video, and group communication and collaboration tools~\cite{marcon-shaping-2011}.

\edit{
To cope with the complexity of mobile network performance dynamics, developers need to measure  and incorporate mitigation strategies in their applications.
Although not all mobile applications are equally affected by poor network performance, the responsiveness of 
%@@ i would be wary of equating responsiveness to QoE
%@@ mwittie 5/5/15: OK, we're citing our results on the link between application responsiveness and poor network performance, but are not linking it to QoE explicitly. QoE is affect to large extent by responsiveness, but there are other causes for QoE and so probably best to leave out.
network applications such as games, interactive video, and, to a lesser extent, in-car navigation and augmented reality requires low latency, stable bandwidth, or both~\cite{Howard14Cascading}.
To improve application responsiveness when latency is high, developers might redesign application communication protocols and message structures to pack more data in fewer round trips between mobile clients and back-end servers~\cite{YahooCombo}.
Developers might also strategically co-locate back-end servers in areas, or networks, where users tend to experience higher latencies~\cite{AmazonRoute53LBR}.
To counteract the effects of low bandwidth, developers, might reduce the size and/or resolution of images and video, or reduce the frequency of application state updates by using techniques such as CloudFlare's Mirage~\cite{mirageCloudFlare} or Opera's Turbo~\cite{operaTurbo}.
}

To apply such performance adaptation techniques, developers need tools to study the performance of \emph{their} application traffic in mobile networks to \emph{their} application servers. 
\edit{
Such {\em in situ} testing is useful during the application design process to reduce the risk of poor application performance at launch, especially when steep user~base ramp-up is expected.
For example, developers may want to evaluate a cloud server placement strategy and measure message delays across geographic areas to validate whether application transactions meet their needs.
Developers may also want to load-test their back-end infrastructure to ensure responsive service, irrespective of user location, server selection, and server load.
Results from such analysis help mobile developers to design appropriate back-end deployment strategies.
Network testbeds are useful for such early testing, because they save developers the trouble of writing testing code, or deploying dedicated back-end servers in multiple locations.
Even during beta testing an application may not have a sufficiently large or distributed user base to generate statistically significant observations, and so community-maintained testbeds are a good alternative starting point.}

Few testbeds support such realistic experimentation prior to application deployment.
Most mobile network testbeds allow users to measure only upload and download speeds, ping latency, and traceroutes, but do not support prototyping of mobile application traffic, or detect traffic shaping in cellular networks.
An alternative to public testbeds are paid services that evaluate application performance across multiple types of mobile devices.
However, these services currently provide access only to stationary cellular devices, which limits measurement realism in terms of geographic and network diversity~\cite{mobitest:webinterface, 14Perfecto}.

Additionally, while there are published best practices for mobile developers~\cite{Grigorik13High}, there are not many tools to track an application's communication performance throughout its lifetime.
\edit{
For example, developers may want to perform A/B testing to
evaluate whether tweaks to communication protocols, or server endpoints, might improve performance. 
Such studies generally target certain users, networks, or time intervals, and thus require expressive test device selection criteria.
Although A/B testing may be implemented in the application itself, third party application optimization libraries offer an easier, safer, and less disruptive starting point~\cite{15Apptimize, 15Splitforce, 15Optimizely}.
However, currently available A/B testing libraries focus primarily on testing application layout with respect to application adoption, user retention, and in-app revenue, but do not collect performance metrics needed to optimize application network performance.
}

Although server monitoring and reporting tools (e.g.,~\cite{Grigorik13High})
%, for example from Librato
enable logging and monitoring of application performance indicators such as request queue length at a server, they do not support end-to-end network measurement.
%@@ is that the only one? this is a 
%@@ strong claim if you only have one example
%@@ mwittie 5/5/15: Utkarsh, could you find a couple other examples similar to Librato and see if what we said about their shortcomings makes sense?
%@@Utkarsh - They are listed here https://blog.profitbricks.com/top-47-cloud-server-monitoring-sysadmin-tools/, 
Other analytics tools, for example the Google Analytics platform, provide performance measurement from a client perspective, but capture only the timing of the request-response cycle, and not response characteristics, e.g., size, compression, protocol~\cite{14Google}.
%@@ again you're making a strong claim based on one
%@@ tool. how many others "analytics tools" are there?  if you don't know of others, just talk about Google's, don't say "all tools like this"?
%@@ mwittie 5/5/15: ditto
%@@Utkarsh - Here is the list of web analytics tools http://en.wikipedia.org/wiki/List_of_web_analytics_software
Thus, developers need to measure and understand application performance in a realistic network environment before and after deployment, particularly as data needs and application requirements evolve.

\subsection{Researchers' View of Network Performance}

The research community has produced several testbeds that offer significant flexibility to execute a variety of network experiments~\cite{goel:mitatepaper, seattle:platform, Nandugudi:phonelabpaper, portolan, netsense, livelabs, mobiperf:homepage, rahul:windriderpaper, aims:myspeedtest, aims:mobilyzer, ndt:mlab, Kreibich:2010:NIE:1879141.1879173, sciwinet:home, phantomnet}.
Yet, the availability of these testbeds and knowledge of how to use them often remains limited by practical barriers to collaboration across research groups.
Researchers may need to set up their own infrastructure for data collection~\cite{Sanchez13Dasu:}, obtain Institutional Review Board~(IRB) approvals~\cite{Nandugudi:phonelabpaper}, or revive code that is no longer maintained~\cite{windrider:mlab, Gummadi02King:}.
Even when maintainers of a given testbed help to set up experiments, communication rounds take time, especially when software modifications are needed.
As a result, researchers often decide it is more expedient to develop new tools, even when it duplicates others' efforts and achieves only a small scale evaluation~\cite{Nandugudi:phonelabpaper,livelabs}.

%@@ mwittie: our purpose for the paragraph below is to prevent reviewers from disputing the conclusion from the previous paragraph if they do think that existing tool repositories are doing a good job at preventing researchers from reinventing the wheel. Thoughts?
Several organizations are working to lower the barrier to entry and promote concerted development of network measurement tools.
For example, M-Lab maintains a repository of measurement tools, including MobiPerf, WindRider, and NDT~(Mobile client), discussed in sections~\ref{sec:mobiperf},~\ref{sec:windrider},~\ref{sec:ndt} respectively~\cite{14M-LabTests}.
One of M-Lab's goals is for new tools to leverage existing code base, for example the Mobilizer library~\cite{aims:mobilyzer}.
M-Lab also supports the development of common ethical guidelines for network measurement data collection~\cite{Zevenberger13Ethical}.
% \todo{mwittie 3/14/15: kc, do you want to mention anything about CAIDA? I think putting on the AIMS workshop is a big step towards getting the community to work together.}
However, the continued flow of proposals for new, independently deployed cellular tools (five in 2014~\cite{phantomnet, livelabs, aims:mobilyzer, Rula14ALICE, sciwinet:home}, eight in 2013~\cite{goel:mitatepaper, seattle:platform, Nandugudi:phonelabpaper, opensignal:homepage, netsense, portolan, fccapp:appintro, ndt:mlab}, three in 2012~\cite{aims:myspeedtest, mobitest:webinterface, radioopt:homepage}, one in 2011~\cite{mobiperf:homepage}, one in 2010~\cite{Kreibich:2010:NIE:1879141.1879173}, and two in 2009~\cite{windrider:northwestern, speedtest:homepage}) suggests that more needs to be done to improve collaboration among different research groups.

\edit{
The research community has also worked to decrease the need for and the cost of redundant experimentation and created several repositories of wireless network measurement data~\cite{15CRAWDAD, 15M-Lab, 09UMass}.
While data repositories facilitate reproducibility of research results, they have their limitations.
For example, to study current phenomena, such as changes in network traffic management policies expected after new FCC Net Neutrality regulations~\cite{Brodkin15FCC}, researchers need new measurement data quickly, rather than waiting for a new dataset to be released after another group's publication.
Additionally, data in repositories may be obfuscated, suitable for one experiment, but lacking in sufficient detail for another, or may be difficult to correlate when multiple datasets are collected at different times or under different conditions.
For these reasons, live testbeds and measurement tools form a critical foundation of innovative research and education environments.}

\edit{
\subsection{Network Operators' View of Network Performance}

In addition to their operational monitoring of cellular network performance from base stations and other network elements, network operators are also interested in end-to-end network measurement from the device's perspective, to provide responsive and reliable service at reasonable operating cost, including the cost of fielding customer support calls.	
Network operators also want to simplify and speed up the deployment of new access technologies and over-the-top services.
A key element in these processes is the ability to troubleshoot network performance issues without affecting base station throughput.

However, industry insiders describe troubleshooting cellular networks as ``an art with few scientific principles.''
To increase their insight into end-to-end network performance and network factors that may affect it, e.g, received signal strength, many network operators have deployed Carrier~IQ on handsets in their networks~\cite{Vijay11ATT, CarrierIQlVodafone}, and then faced customer backlash~\cite{Peckham11Carrier, 11secret} due to this application's approach (or lack thereof) to user privacy protection.
%However, because of controversy over Carrier~IQ user privacy violations, network operators faced customer backlash~\cite{Peckham11Carrier, 11secret}.
Although network operators continue to use Carrier~IQ, users continue to uninstall it on rooted phones~\cite{LookoutLabs13CarrierIQ, sn70714ATT}.
As a result network operators, like ATT, are looking for new methods to monitor and troubleshoot user network performance that can match the scale and efficiency of embedded end-host monitoring provided by Carrier~IQ~\cite{Gopalakrishnan15SDN}.
%@@ need cite for the above, if true. (ATT looking for new methods) 
%@@ mwittie 5/5/15: This is from Vijay Gopalakrishnan's talk at the PhantomNet workshop. We're trying to get a reference. We're also reaching out to Dina from Telefonica to see if she can add to the discussion.
}

\subsection{Regulators' View of Network Performance}

Finally, regulators need monitoring tools to inform their understanding of availability, reliability, and performance of mobile networks over time.
Constrained network performance and delayed upgrades to next generation technologies, e.g., 4G, have long been seen as stifling innovation in the US~\cite{Osborne14state, J.D.Power13Overall}.
Further, traffic shaping mechanisms and anti-competitive behavior by some network providers impede deployment of new services~\cite{nn:verizontoll1, nn:verizontoll2, contentprovider:loser, gamers:netneutrality, claims:isp2, claims:isp3, isp:profit, ISP:askingmoney1, mobilecarriers:chargeperpage}.
Even developers of popular measurement tools struggle to create incentives for longitudinal and widespread measurement~\cite{fccapp:appintro}.
A few tools that have gained traction with users rely on user-initiated network tests, which limits measurement frequency and representativeness~\cite{mobiperf:homepage, Kreibich:2010:NIE:1879141.1879173}.

\subsection{Shared Challenges}

Developers, researchers, network operators, and regulators face the same challenges in deploying end-to-end mobile measurement tools: incentivizing a statistically significant sample of users to install and execute the tool; protecting those users' resources from abuse; and preserving user privacy. 

To motivate user participation, testbed designers have used schemes such as bundling testbed code with other functionality~\cite{Sanchez13Dasu:}, offering free devices~\cite{Nandugudi:phonelabpaper}, press coverage~\cite{fccapp:appintro, Kreibich:2010:NIE:1879141.1879173}, or simply appealing to user altruism and curiosity~\cite{fccapp:appintro}.
These approaches result in either a narrowly focused user base or short-lived deployments, both of which limit testbed utility. 
\looseness-1

The second challenge is how to protect contributed testbed resources from abuse.  
Some peer-to-peer systems have used tit-for-tat mechanisms to ensure fair resource sharing~\cite{Cohen03Incentives}, but mobile network measurement testbeds thus far rely on user altruism on the one hand and conscientiousness on the other~\cite{fccapp:appintro, mobiperf:homepage, rahul:windriderpaper, aims:myspeedtest}.
Scaling and sustaining measurement testbeds over the long term will require more rigorous resource protection methods in existing tools.

Finally, a testbed should isolate personally identifiable information from experimental data collected on a mobile device.
Measurement tools discussed in this paper offer a range of solutions to maintain this separation.
Google has supported the development of a proposed set of ethical guidelines for the design of mobile-based network measurement tools~\cite{Zevenberger13Ethical}.
These guidelines have informed the design of some tools, specifically MITATE and Mobiperf, but the disparate legal frameworks for user privacy around the world make it difficult to create conformant tools for the global mobile Internet~\cite{14AIMS}.

%% file: testbeds.tex
\edit{
\section{\secTestbeds}
\label{sec:testbeds}

Mobile application developers need to know how well a network can deliver their application content.
Custom network experiments that emulate communication protocols of their applications create performance profiles in different network settings to inform application design.
End-to-end systems that support such functionality need to balance the flexibility of their feature set against potential abuse of contributed user resources and threats to user privacy.
We divide systems according to how they resolve this conflict for new experiments from external researchers into uncurated and curated approaches.
}

\edit{
\subsection{Uncurated Network Testbeds}

Uncurated network testbeds allow users immediate access upon registration.
Users experiments and changes to these experiments do not need to go through an approval process. 
Although their open nature allows these platforms to scale, they are limited in the type of personal information they collect without going through an Institutional Review Board~(IRB) approval process.
}

\subsubsection{MITATE}
\label{sec:testbeds:mitate}

Mobile Internet Testbed for Application Traffic Experimentation~(MITATE), developed at Montana State University~(MSU) in April 2013, enables experimentation with mobile application traffic in live mobile networks~\cite{goel:mitatepaper}.
Experiments execute on user-volunteered devices that meet specified criteria, such as signal strength, geographic location, or network provider. 
Developers can use MITATE to evaluate the performance of mobile application communications under a wide range of conditions before their applications are deployed, or even fully developed. 
MITATE supports configurable active network measurements to detect network traffic shaping
by ISPs, and integration with other tools, for example CPLEX to explore protocol configuration tradeoffs through parameter search and optimization~\cite{IBM13CPLEX}. 

\sss{Functionality} MITATE supports active network measurements on mobile devices.
MITATE experiments are configured through XML files that describe the content of experiment data transfers, transport layer protocols, network endpoints, and timing.
An XML configuration also describes criteria that volunteered devices must meet to execute an experiment, such as network type (cellular or Wi-Fi), signal strength, geographic location, network carrier, minimum battery power, and device model.
To ensure that experiments are defined correctly, MITATE servers validate new XML configuration files against an XML schema definition~(XSD).
Users interact with \mbox{MITATE} through an API that allows upload of XML configuration files and download of collected data.

Each mobile device polls a central MITATE server at MSU for new experiments whose criteria matches that device's capabilities.
Devices download static traffic definitions that specify what traffic to exchange between the mobile device and back-end servers.
MITATE mobile devices can interact with third party systems, for example DNS and CDN servers, through explicitly configured, well-formed request packets, and by recording reply content and delay.
Although each MITATE experiment is a series of static transmissions, complex logic can be implemented across processing rounds, e.g., DNS lookups and ping transactions require two rounds.
Such an experiment specifies a device ID as a criteria, which allows for the same device to issue DNS lookups in round one and subsequent pings in round two.

\sss{Data Collection} MITATE records the delay of each data transfer as well as metadata such as signal strength, accelerometer readings, and device location.
This delay measurement allows calculation of 42 metrics, including uplink and downlink latency, throughput, jitter, and loss, as well as mobile sensor readings~\cite{mitate:doc}.
For example, an experiment estimates available bandwidth by dividing the size of a large transfer by its duration.
MITATE experiments may also use a series of small transfers to estimate packet round trip time~(RTT), loss, and jitter.
At the start of an experiment, MITATE estimates the clock offsets between a device and each server endpoint, which allows separate measurement of uplink and downlink latency.

Collected data is available for download in the form of SQL insert statements to populate a local instance of a MySQL database for each user.
MITATE allows users to download data only for their own experiments and those whose data is made public.
Aggregate metrics, for example mean latency, are computed through queries to the local database instance.
This design reduces the load on the MITATE database servers and allows users to run arbitrary queries over their experiment data.

\sss{Resource Incentives and Protection} MITATE is a collaborative framework built around incentives for user participation, inspired by BitTorrent's tit-for-tat mechanism~\cite{Cohen03Incentives}. 
MITATE users earn data \emph{credit} by contributing their mobile resources.
Users can then spend credit to run experiments on others' devices.
Earned credit expires after 24~hours to prevent its accumulation and use for large experiments that might overwhelm available system-wide resources at any point in time.

MITATE's credit system encourages ongoing participation and protects contributed resources from abuse.
Users can leverage MITATE resources in direct proportion to how much data they contribute to the system.
MITATE does not rate-limit device transmissions (although users can set monthly data caps and battery limits on their devices), which permits realistic load-testing experiments.
Although distributed denial of service~(DDoS) attacks launched from multiple devices are technically possible in MITATE, they are destined to be short lived, because rapid transmissions from multiple devices will quickly deplete the malicious user's earned credit.

\sss{Privacy Protection} A significant challenge to expanding measurement systems on volunteered personal devices is the threat to user privacy.
To limit the exposure of personally identifiable information, MITATE captures data only from active traffic experiments and does not monitor non-MITATE device traffic.
Collected data is also indexed by virtual device IDs, rather than personally identifiable phone and International Mobile Equipment Identity (IMEI) numbers.

\sss{Remaining Challenges} 
MITATE is still in active development; project goals for the next couple of years include: deployment on M-Lab, support for peer-to-peer transmission between mobiles (important for IoT and gaming experimentation), and iOS device support.

\subsubsection{Seattle} 
\label{sec:seattle} 

The Seattle testbed, originally developed in March 2009 at the University of Washington to support wired host experimentation, now also supports mobile application prototyping~\cite{cappos2009seattle}.
The design goal was to increase the diversity of testbed hardware to provide a more realistic prototyping environment than testbeds relying on dedicated hardware (e.g., PlanetLab, Emulab, or GENI~\cite{Chun03PlanetLab:, Siaterlis13Use, 12GENI}).
Seattle runs on volunteered devices in last mile networks, and on institutional servers.
As of 2015, Seattle includes about 800 mobile devices and over 10,000 nodes in total.

\sss{Functionality} Seattle experiments run on sandboxed virtual machines in a pared down implementation of Python called Repy.
Seattle libraries support Repy functions such as data serialization, cryptography, and processing URLs, HTTP messages, and other protocols.
Repy code is pushed to Seattle-registered through an API.
Users can select devices by location and network type~(\mbox{Wi-Fi} or cellular) to which device is connected, but Seattle does not support selection by device travel speed, provider, or model.
Seattle also supports P2P communication among devices.

\sss{Data Collection} Seattle does not collect network performance data by default.
Instead users define their own metrics through experiments implemented in Repy.
Seattle does not provide access to device sensors~\cite{open3gmap:planning}.
although sensor applications can make sensor data available to Repy programs through an API.
The Sensibility testbed is an extension of Seattle, which allows Repy experiments to interact with mobile sensor data, but not to transmit or capture network traffic~\cite{14Sensibility}.

\sss{Resource Incentives and Protection} The Seattle incentive model is based on a tit-for-tat approach, where a user has access to ten volunteered devices for every device she registers with the system.
While this policy makes sense in the wired setting, where devices are not generally restricted by monthly data caps,
users who register wired hosts but experiment with others' mobile devices can deplete the mobile data cap. 
As a mitigating step, by default Seattle limits data transmissions to 10~Kbps, so even if the experiment fully uses that transmission rate, the owner can likely continue using their device.
This limit prevents Seattle experiments from measuring available bandwidth and generating load-testing traffic -- limitations not present in MITATE's credit-based model.

\sss{Privacy Protection} Seattle protects user privacy by allowing experiment code execution only in sandboxed virtual machines, which isolates experiment processes from each other and from non-Seattle processes.

\sss{Limitations} The authors of Seattle list several limitation of the current system, including inability for Seattle nodes to host services on ports below 1024, increase the transmission limit on donated resources, send ICMP traffic due to Repy restrictions, and put a limit on battery drain~\cite{seattle:platform}.

\subsubsection{Emerging Systems}

PhantomNet, being developed at University of Utah, is an emerging testbed based on a network of small-cell base stations connected through a software-defined network~(SDN) backbone~\cite{phantomnet}.
Users will be able to not only experiment with end-to-end services, but also modify backbone traffic forwarding for their experiments.
PhantomNet devices will have dual-radio interfaces, which will allow integration with a reseller network, for example through SciWiNet.
PhantomNet also leverages management tools from other systems, notably Emulab and Seattle.
Currently, PhantomNet remains under development.

\edit{
\subsection{Curated Network Testbeds}

Curated network testbeds vet network experiments prior to deployment.
In particular, vetting involves passive monitoring experiments that collect privacy sensitive data, such as users' traffic, or location history and may need to go through an IRB review.
Other experiments may require changes to the testbed itself and need to be approved by the testbed's developer team~\cite{phonelab:agreement}.
}

\subsubsection{PhoneLab}

PhoneLab is a programmable smartphone testbed, developed at the University at Buffalo in November 2013, to support flexible experimentation intended to emulate application deployment scenarios~\cite{Nandugudi:phonelabpaper, phonelab:homepage}.
PhoneLab experiments are implemented as mobile applications pushed to rooted Android smartphones given to student volunteers at the University at Buffalo.
PhoneLab's model supports long-term, passive experiments that can record network transitions, battery drain, and use of other applications on the device.

\sss{Functionality} PhoneLab experiments are pushed to participants either via the Google Play Store, or separate as over-the-air updates.
PhoneLab can benchmark third-party mobile applications without modifications to their code, which may be required in other testbeds.
PhoneLab mobile applications can run experiments in the background or interactively.
PhoneLab also supports experiments at the OS level, with modifications to the Android runtime system.
Platform experiments are vetted by the PhoneLab development team and go through pre-deployment testing. 
Researchers submit experiments as XML configuration files that specify background experiments to start or stop, log tags to collect, and where to upload collected data.
The PhoneLab Conductor fetches configuration files from PhoneLab servers and pushes them to testbed devices.
%@@ how does the above happen?  are all phonelabs servers having mirror of
%@@ the files? or how do the servers coordinate?
%@@utkarsh - this information is not available on their paper/website.
%@@ mwittie 5/8/15: I don't think this is critical for now.

\sss{Data Collection} PhoneLab data collection relies on the Android logging interface, which gives experiments access to device operational data (such as phone status, battery level, etc.), as well as custom application log data.
All log data is uploaded to the central server when a device is charging.
When their experiment completes, users receive an archive of data that matches experiment tags from all devices that participated in their experiment.

\sss{Resource Incentives and Protection} Unlike MITATE and Seattle, which rely on volunteered devices, PhoneLab provides phones with discounted data plans to its participants.
In spite of this incentive scheme, the PhoneLab team has faced significant participant attrition, with only 43 of 191 volunteers continuing after the first year~\cite{Nandugudi:phonelabpaper}.
PhoneLab limits the number of simultaneously active of experiments on each device to balance device utilization against interference between experiments.

\sss{Privacy Protection} To protect user privacy, experiments submitted to PhoneLab need IRB approval or exemption.
PhoneLab participants choose to participate in a particular experiment after reviewing what information will be collected.
Participants can opt-out of an experiment at any time.

\sss{Limitations} PhoneLab's use of data plan subsidy potentially limits the scalability of the testbed. 
%@@ huh, aren't all testbeds operatored w volunteered labor?
%@@ if anthing this is one of the few that is not volunteered,
%@@ because the participants are paid in cell phone service!
%@@ mwittie 5/8/15: what they do now is subsidize a data plan, so I changed to that
Also if phones are not replaced frequently, testbed hardware will eventually lag behind phone models used by the general public.
Finally, PhoneLab code is not publicly available, which precludes the possibility of private deployments~\cite{phonelab:participate}.

\subsubsection{SciWiNet} Science Wireless Network (SciWiNet), being developed at Clemson University, is a NSF-funded re-seller of network infrastructure, based on Mobile Virtual Network Operator (MVNO) model, which provides the research community with a service on Sprint's cellular network infrastructure (and T-Mobile's infrastructure by late 2014)~\cite{sciwinet:home}. 
SciWiNet supports experimentation over 3G and 4G cellular networks, but without support for SMS, MMS, or voice services.
SciWiNet provides additional infrastructure to the research community in the form of a shared pool of wireless devices (smartphones and USB LTE dongles), a common set of Android applications (WiFi hotspot, VPN tunnels, performance monitoring programs), and a set of wireless network services (VPN tunnel termination, secure database backend, performance monitor servers and backend).
%@@ need to contanct jim to update this information

\sss{Deployment} The SciWiNet project has two proposed project phases and is in phase-I as of September 2014. 
In phase-I, the project aims to determine the potential user community for SciWiNet infrastructure and investigate capabilities that it should support.
In phase-II, the project will develop, deploy and operate the functional SciWiNet network infrastructure based on what was learned in phase-I.
%@@ need to update above. 

\sss{Device support} Since SciWiNet uses Sprint's cellular network as its back-end cellular infrastructure, Sprint maintains a whitelist of mobile devices that are authorized to access SciWiNet's network and therefore eliminates the need to install a SIM card in every mobile device. 
Although SciWiNet records device MAC address, it does not make the device MAC publicly available.
SciWiNet maintains a list of popular devices and blacklisted devices.
iOS devices are excluded because they do not support re-seller networks~\cite{sciwinet:devices}.
SciWiNet helps researchers access testbed resources by providing them with 1-2 mobile devices and a prepaid data plan for a limited time, typically six months.
Alternatively researchers can access SciWiNet from their own devices and SciWiNet covers part of the data usage costs.

\sss{Data Collection} SciWiNet Android app collects the following network measurements over cellular and Wi-Fi networks: throughput for TCP and UDP traffic flows, packet loss, and ping latency. 
It can also detect location-based services such as base station identity, location, and wireless signal strength.

\sss{Resource Incentives and Protection}  Users can login to their account to check their data usage, or data contributed by others to their experiments. 
Data usage is limited by a leaky bucket rate limiter, where a user receives a number of tokens, which he can share among multiple devices. 
Once the data rate is exceeded, the device is temporarily restricted from accessing the SciWiNet network.

\sss{Remaining Challenges} 
As of September 2014, it is unclear how SciWiNet will provide access to its devices and network resources to the research and developer community. 
One possibility is to offer incentives for user participation by providing free or discounted device access.
%@@ needs to be updated.  don't think this paragrpah should be here if this is it. 

\subsubsection{LiveLabs}

LiveLabs, designed at Singapore Management University in February 2014, is a mobile testbed intended to evaluate location-based services, such as commercial promotions to shopping mall customers~\cite{livelabs}.
LiveLabs has been tested on the campus of the Singapore Management University~(SMU) and is currently being deployed at a large shopping mall near SMU campus, Singapore Changi International Airport terminal, and on the Sentosa resort island.
The testbed is available to the three partnering venue operators, but not the general public.

\sss{Functionality} To facilitate evaluation of location-based services,
LiveLabs supports device location discovery in indoor settings as well as characterization of user behavior.
LiveLabs is designed for continual operation, thus the design has focused on low energy usage, for example by allowing multiple experiments to concurrently use sensor readings such as GPS, or WiFi signal strength.
Researchers and participating companies use LiveLabs to evaluate location-based applications, for example  real-time promotions to users at a shopping mall. 
LiveLabs is available for Android and iOS systems.

\sss{Data Collection} Unlike other testbeds discussed in this section, LiveLabs does not collect network performance metrics, but instead focuses on discovering user behavior, by recording device ID and a variety of sensor readings.
The LiveLabs backend then supports higher level functions to detect and record user behavior, such as history of movement, group size, user physical queue length, and activities such as standing, walking, or sitting.
LiveLabs also records information about participating users, such as their nationality.

\sss{Resource Incentives and Protection} LiveLabs has three mechanisms for garnering user participation: rebates on users' monthly data bills;
context-based apps that offer rebates on
specific commercial services in deployment locations~\cite{LiveLabs13Participation}; and a ``lucky draws'' lottery, though details of frequency and prizes are not specified~\cite{livelabs}.

\sss{Privacy Protection} Data collected by LiveLabs has the potential to disclose private user information, such as location, shopping patterns, and nationality.
As such, experiments launched on LiveLabs go through SMU's IRB approval process~\cite{LiveLabs:registration}.
Users are also asked to opt-in to data collection on their devices.

\sss{Limitations} LiveLabs is not designed for mobile network measurement (does not collect network metrics) and so it offers functionality distinct from MITATE, Seattle, and PhoneLab.
At the same time, LiveLabs supports experimentation with new services in the mobile environment similarly to PhoneLab and has attracted participation of 30,000 users through its incentive model and business partnerships.

% \subsubsection{Emerging Systems} 

% In addition to MITATE, Seattle, PhoneLab, SciWiNet, and LiveLabs we are aware of two systems in different stages of planning that will support mobile application prototyping.

% {\em NetSense} is a mobile testbed, designed at the University of Notre~Dame in August 2013, focused on characterising the impact of mobile information technology on social tie creation~\cite{netsense}.
% The project is similar to PhoneLab in that two hundred rooted Nexus 4S Android devices with fully subsidized data plans were given to incoming freshmen.
% NetSense has also experienced high participant attrition, mitigated by subsidized device repair and replacement plans.
% As of August 2013, NetSense is winding down its two-year data collection effort and will not continue as a testbed.

%% file: tools.tex
\edit{
\section{Measurement Tools}
\label{sec:tools}

Mobile network performance characterization requires wide scale and ongoing measurement from a variety of devices across different networks and locations.
Tools in this space, developed by industry, research, and regulatory communities, differ in how they obtain network metrics and how they select devices for measurement.
Although network measurement tools presented in this section are not testbeds, in that they only support a fixed set of experiments, these tools do support long-term and wide-scale network monitoring, which offers important insights to developers, researchers, and regulators.
}

\edit{
\subsection{Standalone Measurement Tools}

Standalone measurement tools are ready-to-deploy solutions with pre-defined network measurement functionality.
The open-source nature of these tools allows other to modify them, although many of the tools offer measurement customization options.
Data collected by these tools is generally, though not always, publicly available.
}

\subsubsection{FCC Speed Test}

The FCC Speed Test app, released in November 2013, was designed to provide insight to regulators and the public on the performance of mobile networks across the United States~\cite{fccapp:appintro}.
Developed in collaboration with SamKnows and major wireless service providers, the free application is available on Google Play Store for Android smartphones~\cite{fccapp:play}.
An iOS version of the application is also slated for release, though limitations of the iOS API prevent collection of some metadata that is collected by the Android version~\cite{Johnston13Measuring, fcc:publicity4}.

\sss{Functionality} At the start of a measurement, the FCC Speed Test app pings available measurement servers to identify the one with lowest round trip time~(RTT) to the mobile device.
The selected server then sends a list of measurement instructions to the mobile device.
If the mobile device is currently using less than 64~Kbps of bandwidth for other tasks, it starts the measurements, otherwise the device postpones measurement until its bandwidth usage drops.

The FCC Speed Test app supports active traffic measurements over four types of connections: single connection HTTP GET and POST, as well as multi-connection GET and POST.
Multi-connection transfers test multithreaded download performance over three parallel downloads of 256KB data chunks.
To measure packet loss and RTT, the FCC Speed Test app exchanges a series of UDP packets with the nearby server.
Following a measurement, the mobile device uploads measurement data and associated metadata to an FCC server.

\sss{Data Collection} The FCC Speed Test app reports upload and download rates, packet loss, and RTTs based on HTTP and UDP transfers.
Packet loss on a path is inferred based on failure to receive a UDP packet on that path within three seconds.
The app records the number of packets sent each hour, the average RTT, total packet loss for performed tests, and throughput in 5-second intervals~\cite{FCC13Measuring}.
The app also collects device-related as well as network metadata, including
signal strength reported by the device, connection type (3G/4G/Wi-Fi), location and ID of cell towers, GPS location, device model, OS version, network country code, SIM's operator ID, SIM's country code, network carrier, phone type (GSM/CDMA), and the device's roaming status.

\sss{Resource Incentives and Protection} To build nationwide measurement capacity the FCC Speed Test app relies on  user curiosity about their network performance.
Instrumental to the app's popularity and success was a press campaign~\cite{fcc:publicity1, fcc:publicity2, fcc:publicity3, fcc:publicity4, fcc:publicity5}, which was followed by application installation and measurements from more than 50,000 devices in about 1.5 years.
These numbers have declined over the life of the system, so the effectiveness of a publicity-driven approach to support long-term network monitoring remains to be seen.

\sss{Privacy Protection} The FCC app collects measurement data on the mobile device in the application sandbox, as opposed to through the standard Android logging interface, so data is not visible to other applications.
The collected data are uploaded to FCC servers over encrypted connections. 
Once the data are uploaded, or become stale, they are automatically deleted from the application's sandbox storage. 
The FCC Speed Test app does not collect personally identifiable information, such as phone number or IMEI~\cite{fccapp:privacy}.

\sss{Limitations} The FCC Speed Test app executes only experiments configured by the FCC, i.e., it does not support custom network measurement.
As of October 2014, the configured tests do not detect traffic shaping in mobile networks, which is of increasing interest to regulators and the general public~\cite{nn:verizontoll1, nn:verizontoll2, contentprovider:loser, gamers:netneutrality, claims:isp2, claims:isp3, isp:profit, ISP:askingmoney1, mobilecarriers:chargeperpage}.
With respect to resources used on the device, the FCC application runs at startup and prevents the phone from sleeping, which can drain the phone battery.

\subsubsection{WindRider} 
\label{sec:windrider}

Content-based traffic discrimination has recently been considered a threat to mobile application performance~\cite{nn:verizontoll1, nn:verizontoll2, contentprovider:loser, gamers:netneutrality, claims:isp2, claims:isp3, isp:profit, ISP:askingmoney1, mobilecarriers:chargeperpage}. 
WindRider, a measurement tool developed in 2009 at Northwestern University, detects application and service-based traffic discrimination by mobile ISPs~\cite{rahul:windriderpaper}.

\sss{Functionality} WindRider supports active and passive measurement of traffic shaping~\cite{windrider:northwestern}.
Active measurements exchange traffic between a user's mobile devices and a randomly chosen M-Lab server.
The mobile device initiates a series of uploads and downloads and records their observed performance.
To detect port-based traffic shaping, WindRider compares delay of identical transfers to different ports on \mbox{M-Lab} servers.
Passive measurements record packet latency to well-known web servers during normal user browsing activity.
To detect content-based traffic shaping, WindRider compares the observed packet delay to that reported by other devices in different carrier networks and locations to the same destinations.
Active measurement results are stored on M-Lab servers, while passive measurement data, collected with user permission, are stored on WindRider servers.

\sss{Data Collection} The WindRider mobile application collects experiment-related data such as connection start time, connection establishment time, connection finish time, and number of inbound and outbound bytes~\cite{rahul:windriderpaper}.
WindRider also collects metadata such as device IMEI, device location (as ZIP code), network carrier, and browsing history.
WindRider also collects device hardware performance metrics that can help interpret observed traffic delays, such as CPU execution time, virtual memory size, page faults per minute, and other metrics as permitted by the OS API.

\sss{Resource Incentives and Protection} WindRider relies on user curiosity for its network measurements.

\sss{Privacy Protection} WindRider optionally collects device IMEI, which can be linked with a user's browsing history.
To protect user privacy, users can choose whether to make this information available to the application.

\sss{Limitations} 
Although \mbox{WindRider} supports detection of traffic shaping in mobile networks, it has two significant limitations.
First, the measurement traffic is sent only to M-Lab servers, but developers may want to investigate traffic shaping on other paths.
Second, WindRider only detects content-based traffic shaping as discrimination based on traffic sources, i.e., well-known Web servers, rather than type of traffic, for example BitTorrent.

\subsubsection{MySpeedTest}

The MySpeedTest mobile application, launched in June~2012 by Georgia Tech, measures network performance of mobile devices with the goal of observing and explaining patterns of user behavior in mobile ISPs to application developers~\cite{myspeedtest:thesis, Feamster12MySpeedTest}.
Such analysis may allow developers and service providers to tune application performance~\cite{Muckaden13MySpeedTest:}. 
%@@ is there any evidence that service providers are helping application
%@@ developers tune app performance?
%@@ mwittie 5/8/15: this point came from an MS thesis and we added a citation
The MySpeedTest mobile application is available on Google Play and has more than 900 active users from 115 different countries, as of February~2013~\cite{Muckaden13MySpeedTest:}. 
As of April~2013, MySpeedTest is in the process of sharing a subset of their data with Google's \mbox{M-Lab} to help researchers benefit from data collected by each others' experiments~\cite{myspeedtest:thesis}.

\sss{Functionality} MySpeedTest performs passive and active measurements.
Passively, MySpeedTest records the total number of bytes sent and received by each active application since the device booted.
Information such as package name, bytes transmitted and received, application status (active vs. background) helps users know which applications consume the most data and  power, and which applications may affect performance of other applications on the device.

Active measurements include a recurring test to measure TCP uplink and downlink throughput, inter-packet delay, and packet loss. 
MySpeedTest also measures network latency with 40 parallel ICMP pings to five servers in the U.S.~and Europe.
These tests store the minimum, average, and maximum latency to each of the five servers.
The collected data help researchers and developers understand the performance of paths to potential application servers~\cite{Muckaden13MySpeedTest:}.

TCP-based experiments can reduce the bandwidth available to other applications on the device, so MySpeedTest performs TCP-based experiments only on user request, in a single thread for about 20 seconds, and using the maximum-sized packets that will not be fragmented. 
MySpeedTest also gauges streaming data quality by measuring packet loss and jitter of UDP traffic flows. 
MySpeedTest servers 
%@@ which server? an MLab one?  at gatech? how many are there?
%@@ mwittie 5/8/15: I checked with Utkarsh and they don't specify where the server is hosted.
generates a stream of 64-byte UDP packets, transmission at Poisson-sampled intervals, with timestamps and sequence numbers in the payload.
The server sends 500 packets with a data rate less than 1\,Kbps to avoid congestion. 
The client calculates packet loss and jitter from every 10 packets received.  
The client compiles all data collected on mobile device 
into the JSON format and sends it to the server for storage.

\sss{Data Collection} The MySpeedTest mobile application collects experiment-related data such as TCP upload and download throughput, ping latency, UDP jitter, UDP packet loss, and time to acquire a dedicated channel for data transmission~\cite{myspeedtest:thesis}. 
MySpeedTest also collects device level data, such as cellular service provider, Android version, device manufacturer, connection type, radio firmware, hashed phone number, hashed IMEI, software version, SIM card state and serial number, latitude and longitude of base station, network operator ID, CDMA system ID, CDMA network ID, \mbox{Wi-Fi} signal strength, battery technology, status of battery charging, battery health, battery voltage, battery temperature, and device location.

\sss{Resource Incentives and Protection} Similar to the FCC Speed Test app, MySpeedTest relies on user curiosity about their network performance.
MySpeedTest allows users to limit contribution of resources through a monthly data cap.
To protect battery resources, MySpeedTest postpones experiments until the battery is above 5\% and the device is attached to a network.
%@@ does it run in the background?  how often does it check battery and
%@@ network state? 
%@@ mwittie 5/8/15: KC, the level of information we're offering comes from the paper. We can break into the code, or run the application and measure stuff ourselves, but then we would need to explain our experimental methodology, which I think is our of scope for a survey. Also, I think it's more important to say what is it that each system does (i.e. what design angles its developers considered) rather than compiling what might read as a documentation page for each tool.

\sss{Privacy Protection} MySpeedTest collects personally identifiable information (phone number, IMEI, device location), which may expose private information, such as a user's location when a measurement occurred.  

\sss{Limitations} Similar to MobiPerf and WindRider, MySpeedTest provides its users a limited network measurement capability between mobile devices and servers, as opposed to testbeds discussed in Section~\ref{sec:testbeds}.
MySpeedTest does not support transmission of custom traffic, such as tools to
detect traffic-shaping based on content or port.

\subsubsection{Akamai Mobitest} 

Akamai's Mobitest application and Web service, released in March 2012 by Akamai Technologies, measures the performance of mobile Web sites~\cite{mobitest:webinterface}. 
The application uses the WebPageTest framework and is available for Android, iOS, Blackberry based smartphones, 
tablets and simulators~\cite{mobitest:WebPageTest}.

\sss{Functionality} Mobitest platform relies on user participation to install Mobitest software on their mobile devices.
Each Mobitest installation on a device acts as an agent to the WebPageTest framework, where such device executes experiments requested by other users through the Mobitest Web service~\cite{mobitest:blog}.
To measure the page load time on a mobile device, a user enters a URL through the Akamai Mobitest Web interface and selects the mobile device hardware that will perform the download~\cite{mobitest:webinterface}.
Mobile devices running Mobitest periodically poll WebPageTest servers to obtain pending URL download requests entered by Mobitest users.
Each requested URL is then accessed from the default browser on each device over the \mbox{Wi-Fi}, or cellular network, depending on how the device is connected at the time. 

\sss{Data Collection} Akamai Mobitest collects the total time to load a Web page, individual request headers, average Web page size, as well as 
screen shots of the loaded page and optionally video of the loading page~\cite{mobitest:blog}. 
The tool produces waterfall charts of requests and delays, and an HTTP archive~(HAR) file~\cite{mobitest:har, mobitest:register}.
The collected data helps researchers and developers gain insight into the responsiveness of Web servers and browser rendering of different site implementations~\cite{Piatek14Measurement}.
Mobitest allows users to reuse previously collected measurements by linking 
them to user accounts on Akamai Mobitest's site.

\sss{Resource Incentives and Protection and Privacy} The Akamai Mobitest app allows application developers to set the frequency at which pending experiments are downloaded from WebPageTest servers to be executed on their mobile devices. 
Additionally, Akamai Mobitest allows users to control device resource utilization through a number of configuration options.
Specifically, users can set whether the app should poll for new experiments after restart, whether to restart the app after every experiment, whether to capture network traffic, and the frequency at which screenshots for loading pages are taken~\cite{mobitest:protection}. 
\looseness-1

\sss{Limitations} 
Akamai Mobitest evaluates the webpage load time on mobile devices, but does not allow more general experiments with non-browser-based application traffic, including how to characterize traffic shaping of non-Web traffic.
The WebPageTest framework requires rooted phones, which limits the tool's applicability outside of dedicated test farms.\looseness -1

\edit{
\subsubsection{RILAnalyzer}

RILAnalyzer, developed by the University of Cambridge and Telefonica in October 2013, is a client-side tool for monitoring of the mobile network control plane as well as the data plane~\cite{Vallina-Rodriguez13RILAnalyzer, Aucinas13Staying}.
The application is available for rooted Android devices with Intel/Infineon XGold chipsets, which include the popular Samsung Galaxy S2/S3, Note 2, and Nexus devices.

\sss{Functionality} RILAnalyzer's focus is on discovering the promotions and demotions between the Radio Resource Control~(RRC) states \texttt{IDLE}~(no connection), \texttt{CELL\_DCH}~(dedicated communication channel), \texttt{CELL\_FACH}~(shared communication channel), and \texttt{CELL\_PCH}~(shared paging channel).
Transitions between these states are triggered by control messages from the Radio Network Controller~(RNC), which may themselves become a communication bottleneck~\cite{Vallina-Rodriguez13RILAnalyzer}.
As mobile devices consume different levels of energy in each of the RRC states, the devices themselves may use \emph{Fast Dormancy} to reduce \emph{tail-energy} and demote to lower energy states faster than through vendor and operator dependent timeouts~\cite{Labs11Understanding}.

RILAnalyzer implements a background tool that polls the device Radio Interface Layer~(RIL) Daemon every second for the current RRC state.
RILAnalyzer then obtains data plane network and transport headers using NetworkLog~\cite{11NetNetwork} to identify applications active during each RRC state.

\sss{Data Collection} RILAnalyzer collects RRC states at one second intervals, headers and timestamps of outgoing TCP and UDP packets from NetworkLog as reported by the Linux kernel.

\sss{Resource and Privacy Protection} RILAnalyzer is intended for small scale studies on dedicated devices, or devices operated by expert users~\cite{Vallina-Rodriguez13RILAnalyzer}.
As such the tool's design has not made provisions to attract users with incentives, or to allow them to set limits on resource usage.

\sss{Limitations} RILAnalyzer is restricted to rooted phones on the Intel/Infineon XGold chipset.
Although the authors of RILAnalyzer intend the tool for small scale studies, the specificity of hardware and overhead of reverse engineering RIL Daemon \texttt{OemCommands} commands does preclude large scale studies on diverse mobile hardware.
RILAnalizer also puts a noticeable load on the CPU~($\sim\!\!10\%$), memory~($<\!\!42\%$), and storage~(with packet logs), which may limit the willingness of volunteers to run the tool on their phones.
}

\edit{
\subsection{Libraries for Mobile Network Measurement}

Libraries for mobile network measurement may be embedded in other applications to add network measurement functionality. 
This approach is potentially easier to adopt by Developers than extending open-source code of a standalone measurement tool.
As in the case of Mobilizer, a library may also form a basis of a measurement tool, i.e. the current version of MobiPerf.
}

\subsubsection{MobiPerf} 
\label{sec:mobiperf}
The MobiPerf mobile application was developed as a collaboration of University of Michigan, Northeastern University, University of Washington, and Google's
M-Lab to measure network performance and diagnose problems with application content delivery on mobile devices~\cite{mobiperf:homepage}. 
To allow the community to understand the impact of collected data across geographic locations, network carriers, and devices, MobiPerf allows a comparative study of past network measurements made by different users, 
but prevents users from running similar measurements to limit 
contention for testbed resources.
New measurements are executed only if a query for previously collected data comes back empty.
%@@ kc2mike: i'm confused: why would two measurements at different times be expected to be substitutable for one another? 
%@@ mwittie: from my discussions with David it seems he thinks they might be close enough. I also don't think one measurement is a good substitute for another. Maybe good enough for some applications? Anyway, that's what they do.
The latest version of MobiPerf, released in August 2014, is based on Mobilizer -- an open-source Android library for network measurement announced at AIMS 2014~\cite{aims:mobilyzer}.

\sss{Functionality} MobiPerf supports several types of network performance measurement, which can execute serially or in parallel~\cite{mobiperf:details}.
Mobilyzer provides measurement isolation (only one experiment is active at a time), which avoids bandwidth contention and radio power state transitions across experiments.
To measure throughput, Mobiperf transmits random data to and from a nearby M-Lab server for 16 seconds and computes uplink and downlink throughput from packet traces.

MobiPerf supports latency measurements on both IPv4 and IPv6 network paths, using ICMP ping when available, with fallback to a Java ping implementation and latency estimates from three-way TCP handshakes in HTTP transfers. Mobiperf measures the delay of DNS lookups using the default DNS server configured for the device, which limits the ability to measure performance of third-party open DNS infrastructure.

MobiPerf also supports measurement of uplink and downlink UDP packet loss, out-of-order delivery, and variation of one-way latency. 
To obtain these metrics on the uplink, a client device sends a group of UDP packets to a nearby \mbox{M-Lab} server, where the server calculates network metrics from packet arrival time and order.
The same transmission repeats from server to client to calculate downlink metrics.

MobiPerf performs more complex measurements to discover fine-grained network policies and their effect on data plane performance. 
For example, MobiPerf measures radio resource control~(RRC) state information of cellular networks to estimate the impact on packet latency~\cite{Rosen14Discovering}.
Finally, MobiPerf measurements can execute in the background to support long-term monitoring of network performance.

\sss{Data Collection} Similar to other measurement tools, the MobiPerf application collects performance data such as TCP uplink and download throughput, HTTP download latency and throughput, traceroutes, path latency, and DNS lookup delay.
 Researchers and vendors may want to know how variation in mobile hardware affects application performance, so MobiPerf collects device-related data such as manufacturer, model, operating system version, Android API level, carrier, salted hash of device IMEI, coarse-grained cell ID location information, cell tower ID and signal strength, Location Area Code (LAC), local IP address, IP address seen by the remote server, GPS coordinates, ports blocked by cellular provider and network connection type (HSPA/LTE)~\cite{mobiperf:datacollection, mobiperf:ppt}.

\sss{Resource Incentives and Protection} MobiPerf relies on user curiosity to support measurement, and users can limit the resources they contribute.
Specifically, measurements do not execute when the device battery consumption, or MobiPerf application monthly data usage, exceed user-set thresholds.

\sss{Privacy Protection} MobiPerf currently records the users' e-mail address, if they choose to provide one, to access their historical measurement results. 
This information is secured by Google's account authentication mechanisms and is not made publicly available. 
To minimize any risk of exposing this potentially personally identifiable information, future versions of MobiPerf will store a salted hash of users' e-mail addresses instead.

\sss{Limitations} MobiPerf allows users to choose from only predefined measurements, which limits the tool flexibility.
For example, MobiPerf does not support transfers of custom content on arbitrary ports to detect network traffic shaping.
%@@ but how many of the tools do support this?  
%given how many tools you have added that don't support this,
%@@ this starts to look like a weird thing to put in here.
%@@ You don't mention iot in all the other tool descritions? 
%@@Utkarsh - should we include iot related experiments in limitations?
%% mwittie 5/8/15: KC, Utkarsh, we mention IoT as an extension of MITATE, but I think it would be to harsh to other tools to list lack of explicit support for IoT as a shortcoming. Utkarsh also changed 'Research Challenges' to 'Limitations' to describe how each testbed, tool, and service does not provide features that other platforms in the same category do. In other words we are comparing limitations of one tool to other tools with respect to things that 'tools' do.

\subsubsection{ALICE}

A Lightweight Interface for Controlled Experiments~(\mbox{ALICE}) is a programmable network measurement library for Android devices developed by John Rula~\textit{et~al.} at Northwestern University~\cite{Rula14ALICE}.
ALICE extends Dasu, a rule-based network testbed built as an add-on to the Vuze BitTorrent client~\cite{Sanchez13Dasu:}, by enabling experiment definition in Javascript~\cite{Rula14ALICEa}.

\sss{Functionality} The ALICE measurement library supports active and passive experiments on mobile devices.
\mbox{ALICE} provides a programmable interface for the configuration of active network measurements, such as DNS resolution, ping, and iPerf.
Tests can execute sequentially or in parallel.
Although the sequence of tests and value passing between them is organized through a Javascript experiment definition, ALICE does not support custom traffic generation, and so is primarily a network measurement library.
For serially scheduled experiments, ALICE allows one experiment on a device at a time; for parallel execution, ALICE allows a limited number of experiments to run at the same time -- new experiments scheduled for a given device enter a queue until the device becomes available.
ALICE chooses its test devices based on user-specified time of day, network provider, and network type~(Wi-Fi/Cellular).

\sss{Data Collection} ALICE collects device location, radio signal strength~(WiFi and cellular), WiFi access point name, device hardware address, IP address on each network interface, and number of bytes sent and received by other applications on the device.
ALICE also collects performance metrics, including HTTP GET request time, DNS lookup time, ping times, available bandwidth.
%@@ how is available bandwidth measured?
%@@ this is very tricky in wired space.. 
%@@utkarsh - we don't have information on how they measure available bandwidth
ALICE records network diagnostic information provided by traceroute and NDT (Section 3.3.1).

\sss{Resource Incentives and Protection} 
As of September 2014, ALICE has been included in three different applications developed at Northwestern University and available through the Google Play store: Namehelp Mobile\footnote{Namehelp Mobile measures the DNS performance of Cellular ISPs and public DNS resolvers, including of CDN replicas~\cite{namehelp:googleplay}}, Application Time~(AppT)\footnote{Application Time allows users to track their application usage on their mobile device~\cite{AppT:googleplay}.}, and \mbox{NU Signals v2}\footnote{NU Signlas allows users to diagnose Wi-Fi problems~\cite{NUSignals:googleplay}.}.
The Northwestern team's deployment model of growing the tool through application deployments allows ALICE to benefit from popularity spikes of new applications.
To protect device resources, developers can set quotas for bandwidth usage of individual measurements.

\sss{Privacy Protection} ALICE records hardware addresses of available network interfaces, 
which are unique to each device. 
In combination with the ability to record sent and received traffic payload of other applications, for example location reporting, ALICE creates a potential for privacy exposure, if user location, or other private data, is correlated to unique device ID.

\sss{Remaining Challenges} Currently ALICE does not support repeatable experiments on the same device, or set of devices, through device selection criteria.
ALICE also does not support peer-to-peer experiments, or custom traffic transmissions, which limits the tool's support for application prototyping.

%% file: services.tex
\edit{
\section{Measurement Services}
\label{sec:services}

In addition to testbeds and tools there are many closed-source, proprietary measurement services for mobile networks. 
We divide these network monitoring and network discovery and diagnosis.
%@@Utkarsh - i changed "network discovery and diagnosis" to just "network discovery". In my opinion network discovery could also indicate network diagnosis. Basically, if you discovered something in the network, you have indirectly diagnosed it.
%If you dont agree with this, let me know and I will change that in the table and in the section title below.
%% mwittie 5/15/16: yes, please change network discovery back to network discovery and diagnosis
The main goal of these is to collect data and provide insight to users based on their own device, but not necessarily make the data broadly available. 
Still, these services offer valuable insight to developers, researchers, regulators, and network operators able to access the data.
Because the details of how these services are implemented and how they perform measurements is not widely available, we restrict our discussion, with few exceptions, to the commonalities and differences of what data these services collect.
%@@Utkarsh - the above sentence applies to only services that do network monitoring, but not to NDT/Netalyzer that do network discovery.
}

\edit{
\subsection{Network Monitoring}

Google Play Store and Apple App Store offer tens of applications for monitoring of network performance. 
Because of their relative similarity, we restrict our discussion to several popular and representative services.
}

\edit{
\subsubsection{Ookla SpeedTest Mobile}

Ookla's SpeedTest application for mobile devices, released in January 2009, measures the device's network performance over \mbox{Wi-Fi} and cellular links~\cite{speedtest:homepage}.
As of March 2015, the application support measurements against 3479
geographically distributed Ookla servers in about 80\% of world's ISP networks, has over 10 million installations, and has successfully completed over 7 billion user initiated measurements on the Ookla infrastructure~\cite{ookla, speedtest:googleplay}.
Ookla also allows users to host an Ookla server 
To expand the capacity of their measurement infrastructure, Ookla also allows users to host an Ookla server~\cite{speedtest:hosting}.
The application is available for Android, iOS, Windows phone, and Amazon FireOS based smartphones~\cite{speedtest:homepage}.

\sss{Functionality} The application captures the device geographic location 
and uses it to identify a set of five nearby servers.
If the device location is not available from the GPS,
the application uses device's IP address and estimates the device's location 
using MaxMind's~(approximate) IP-to-location database~\cite{speedtest:maxmind, speedtest:geoip, maxmind}.
After identifying a pool of five nearby servers, the application sends a \texttt{hello} message to all five servers and selects the measurement server from the first received reply~\cite{speedtest:serverselection}.
Users may also select a specific server based on criteria
such as hosting ISP, distance from user, and city name.
 
This SpeedTest uses HTTP fetches of small files to measure round-trip time and
compute uplink and downlink throughput~\cite{speedtest:tests}.
To measure the ping latency, the application sends several HTTP requests and records the time when app receives responses from the server~\cite{speedtest:tests}.
Ookla SpeedTest uses the computed connection throughput to estimate how much data it can download from the server within 10 seconds, and then uses up to four HTTP threads on a single persistent connection to download the estimated amount of data. 
% It collects throughput values 30 times a second and aggregates into 20~time buckets after the experiment completes.
%@@ 30 times a second?  why? why 20 time buckets?
%% mwittie 5/8/15: We don't know why, that's just their configuration. Frankly, I think that's TMI. I'm commenting it out.
To eliminate any influence on the throughput results from protocol overhead, buffering time on the device, CPU usage, the application first discards the fastest and slowest 10\% of throughput values as outliers before computing the average throughput.
%@@ why would discarding the fastest eliminate influence of these things?
%% mwittie 5/8/15: not sure that is their rationale for this. We suspect is that they want to show steady state, and not the lower throughput during TCP ramp up. So maybe to be fair they are also discarding the fastest transfers as not part of steady state to be included in the average. Who knows. Their paper doesn't say. We include these details so that readers know something non-standard is implemented in metric calculations.
The application then discards the slowest 20\% of throughput values to prevent  results from being influenced by TCP slow-start.
Finally, the application calculates the downlink throughput for the experiment based on the average of the remaining throughput values.
The uplink throughput test is similar to downlink throughput test.

\sss{Data Collection} Ookla's SpeedTest mobile application collects device location~(GPS and network-based), radio signal strength, device~ID, device phone number, call status and remote phone number of an active call, names of devices on connected \mbox{Wi-Fi} network, local and public IP addresses, time at which the experiment was conducted, round-trip time, upload and download throughput, and connected network type~(\mbox{Wi-Fi} or cellular).
Ookla supports another application, {\em PingTest}, that collects network jitter and packet loss, to understand the suitability of the user's network for services such as VoIP audio, video streaming, and online gaming~\cite{pingtest}.

\sss{Resource Incentives and Protection} The application limits the number of HTTP threads to two when the observed throughput is less than 4\,Mbps, otherwise the it uses four threads for throughput experiments.

\sss{Privacy Protection} The SpeedTest mobile application collects personally identifiable information (phone number, device ID, and device location), which may expose private information, such as a user's location when a measurement occurred.
Users may delete previously collected data, or leave it on Ookla servers to compare with new data collected at a later time to discover changes in network performance over time.

\sss{Limitations} As of March 2015, the SpeedTest mobile application lacks a programming interface to allow users to automate and schedule experiments.
Although, Ookla allows users to host SpeedTest experiments on their Web servers for in-house testing, via SpeedTest Mini, however, as of March 2015, the ability to run measurement against such servers is not supported on Speedtest's mobile application and is only supported with the Web version of Ookla SpeedTest~\cite{speedtestmini}.
% removed too much detail, and does't seem fair, it's not considered a challenge by them is it?
%@@Utkarsh - I think it is an important point to mention, especially because I renamed the section to limitations.
The algorithm used by the application to measure the round-trip time relies on the time it takes to receive an HTTP response, which may include the time request spent in transport queue and application processing at the server.
Finally, the application does not support detection of traffic shaping.
}

\edit{
\subsubsection{RadioOpt Traffic Monitor}

The RadioOpt Traffic Monitor mobile application, released in April 2012 by RadioOpt GmbH, allows users to understand the performance, reliability, and utilization of their wireless and cellular networks~\cite{radioopt:homepage}.
Based on the information collected about the network, the application allows users to compare the performance of their wireless networks with other users
in the same geographic region.
The application is available for Android, iOS~(iOS~7.0 or later), Blackberry, and Windows-based smartphones~\cite{radioopt:googleplay, radioopt:download}.
As of March 2015, the application was installed over a million times.

\sss{Functionality} The RadioOpt Traffic Monitor mobile application uses CacheFly's CDN infrastructure. To identify a nearby server, 
the application sends a DNS query to the device's default DNS server for a CacheFly CDN domain name~(\texttt{\url{cdn2.speedtestsdk.com}}).
CacheFly uses TCP-anycast to direct users to the nearest CDN replicas~\cite{cachefly}. 
Next, the application sequentially initiates downlink and uplink throughput tests to the selected server.
To measure throughput, the application estimates the appropriate size of the data to exchange between the device and the server, similar to Ookla SpeedTest.

To measure latency, the application sends 15 ICMP ping requests to the server and records the time of each request/response pair.
To measure the time to load a webpage on user's network, the application sends three HTTP GET requests and records the time to download the complete webpage, and other web objects such as CSS, image, JavaScript files embedded into the page.
%@@ all this seems redundant with the other tools, and yet not
%@@usefull compared, just looks like cutting and pasting of different
%@@ text from different papers rather than extracting similarites and
%@@difference.  can you find a way to shorten into some canonical 
%@@measurements/metrics and just refer to them without explaning?

\sss{Data Collection} The application computes parameters from measurement data such as the minimum, average, maximum ping latency to a nearby CacheFly server along with the standard deviation in latency and throughput, the amount of data uploaded and downloaded for the throughput tests, download time of a hosted web page and the web page size.
The application also collects device-related information such as its location~(including accuracy and device travel speed), web bookmarks and browsing history, names of devices connected to the same \mbox{Wi-Fi} network, signal strengths at different locations, number of SMSes sent and received, and incoming and outgoing voice minutes, device model and manufacturer, OS or firmware version, current time on the device, the time when the device was last rebooted, cellular access technology~(2G/3G/4G), and network country code.
%@@ why so much data? how is it used?
%@@Utkarsh - we dont have information on how they use this data. 

The application also collects information specific to applications on the device such as their names, duration of usage, cellular and \mbox{Wi-Fi} data consumption~(only on Android based smartphones), memory consumption, traffic (per application) sent and received on the device over cellular and \mbox{Wi-Fi} networks, application type~(OS service or background), and software packages used by the application.

The application also collects device battery-specific information such as the battery state and charge remaining, voltage, temperature, technology, and charging state.
Finally, the application collects \mbox{Wi-Fi} network related information such as the signal strength~(latest, minimum, and maximum),
 network SSID and BSSIDs, the MAC address of the client, IP address of the client, and client-to-router link bandwidth.

\sss{Resource Incentives and Protection} RadioOpt relies on user curiosity to understand the performance of their own wireless and cellular networks.
The app allows users to configure a monthly/weekly/daily cellular data cap, monitor their monthly data traffic, SMSes received and sent, and voice minutes, and configure alerts when data, SMS, or voice minutes reach a threshold.

\sss{Privacy Protection} The application may discover user behavior since it collects information such as the user's Web browsing history, bookmarks, applications installed and their duration of usage, among others.
However, any personally identifiable data collected by RadioOpt mobile application is not shared with RadioOpt servers without the user's consent.

\sss{Limitations} RadioOpt does not allow its users to understand whether their cellular ISPs are discriminating one traffic over the other.
Further, the application does not support measurement experiments to be run against an arbitrary server.
}

\edit{
\subsubsection{OpenSignal}

The OpenSignal mobile application, released in March 2013 by OpenSignal, Inc., allows users to compare the quality and coverage of their cellular networks~(on a Google Map's developer widget~\cite{googlemaps}) in different geographic areas and with other cellular networks available in the area~\cite{opensignal:homepage}.
The application rates for how well Web, Video, and VoIP based applications are likely to perform on the current cellular network. 
The application assist users to also find publicly available free and paid \mbox{Wi-Fi} \mbox{hot-spots}, and the walking directions for higher signal strength.
The application has over 10 million installations and is available for Android and iOS based smartphones~\cite{opensignal:googleplay, opensignal:applestore}.
As of March 2015, the application has garnered over 900,000 users and has performed several network measurements to collect information for over 800,000 cellular towers, 825 cellular networks, over 5B cellular signal readings, and over 1B \mbox{Wi-Fi} access points available in different countries~\cite{opensignal:homepage}.

\sss{Functionality} The OpenSignal mobile application supports several active and passive measurements to measure ping latency, download and upload throughput.
The application performs periodic passive measurements, and publishes them to an OpenSignal server~\cite{cainey2014modelling}.
Before starting any measurement test, the application sends the device ID, OS, Android API version, and BSSIDs of nearby wireless networks to an OpenSignal server.
Next, to measure latency, the application sends 3 HTTP HEAD requests to \mbox{\texttt{www.google.com}}~\cite{opensignal:latencytest1, opensignal:latencytest2}.
The application then records the time to receive the time to get the response for each request, followed by calculating the average of the three latency values.

To measure the download throughput, the application sends eight concurrent HTTP GET requests to download files of size 108\,Mb each, from a CloudFront's CDN replica~\cite{cainey2014modelling}.
The download throughput test is performed for a fixed amount of time after which the application computes the average throughput.
To measure the upload throughput, the application sends several concurrent HTTPS POST requests to upload several small image files of size 15\,Mb in total, to an Amazon AWS server.

As a part of making the collected data available publicly and to encourage developers, researchers, regulators, and network operators to investigate and address network problems, OpenSignal provides two APIs~\cite{opensignal:blog}.
The first API, known as NetworkStatus, allows developers to get signal strength, upload and download throughput, round trip latency, and network name, network ID, network type~(2G/3G/4G), and network reliability for every measurement within certain distance of a specified geographic coordinate~\cite{opensignal:networkstatus}.
The second API, known as Tower Info, allows developers to get the cell ID, location area code, phone type~(GSM/CDMA), and estimated latitude and longitude of a cellular tower~\cite{opensignal:towerinfo}.
To prevent misuse of their publicly available API, OpenSignal allows a maximum of five API calls every minute and 2000 API calls every month.
%@@ KC: so what researchers are using this data, and how?
%@@Utkarsh - To answer your question, I added a citation to the first line of this paragraph.
%@@ KC: why do researchers need anything else? 
%@@Utkarsh - I am not sure how to answer this.

\sss{Data Collection} The OpenSignal mobile application collects device-related information such as SMS transmission and receipt timestamps, device location, ID, model name, OS, Android API level, IP address, behavior at different battery temperatures~(hot, crashed, slow, fast), duration of OpenSignal sessions on the device, and whether the phone is engaged in a phone call during the measurement.

The application collects network-related information such as the active \mbox{Wi-Fi} SSID, names of devices connected to the \mbox{Wi-Fi}, SSIDs of other avaiable \mbox{Wi-Fi}, connection type~(collected every 15 minutes), signal strength, upload and download throughput, and round trip latency to a Google server.
For devices connected to GSM networks, the application associates cell towers by their cell id and location area code; for CDMA networks, by their Network ID, Base sub-station ID and system ID~\cite{opensignal:googleplay}.
To understand the relationship between signal quality and battery consumption, the OpenSignal application collects the battery level, voltage and temperature~\cite{opensignal:batterinfo}.

\sss{Resource Incentives and Protection}  OpenSignal does not provide any incentives for user participation to run measurement experiments on mobile devices.

\sss{Privacy Protection} Although the application collects information about phone calls and SMS messages, the application never reads them~\cite{opensignal:privacy}.
This is because the application only counts the total number of text messages received and sent from the device.
Further, any personally identifiable information collected by the OpenSignal mobile application is never shared by any third party services~\cite{opensignal:privacy}.
However, OpenSignal does not take any responisbility of any data shared by the user on online social networking websites through the OpenSignal application.
Finally, the application does not put any obligation on the user to share the data collected with OpenSignal.
%@@ but you said opensignal shares all the data earlier.
%@@Utkarsh - only the data that users want to share.

\sss{Limitations} The application does not detect the presence of traffic shaping in ISP networks.
Further, to perform throughput measurement tests, the application requires an exchange of several hundred of megabytes between the mobile device and the server, which may not be suitable for users with low data plans~\cite{cainey2014modelling}.
}

%@@ i am strongly opposed to having the below section
%@@ be anything other that a short paragraph explaining
%@@ what has actually been learned from the tool
%@@ in concrete terms.  if it's only for vodafone to
%@@ make use of the data itself, i don't see why it's
%@@ in this paper except to explain that operators do do
%@@ their own measurements but none of that is available
%@@ to researchers.  by this time i was really frustrated
%@@ with the paper, and just crossed this entire section out,
%@@ after i heavily marked it up on the ipad ;(
%@@ my vote is to reduce it to 2-3 sentences, like most of
%@@ the other tools added to appease reviewer 2 
%@@Utkarsh - So i got vodafone's android apk from someone in EU and i tested the mobile app and found that the application also works for users (in USA) not on vodafone network. So, i believe we should have a section on vodafone netperform, similar to others.

\edit{
\subsubsection{Vodafone NetPerform}

The Vodafone NetPerform mobile application, released in June 2014 by Vodafone Sales and Services Limited, allows users to understand the performance of their cellular network in their region and compare with it with the performance that other users in the same region are experiencing~\cite{vodafone:homepage}.
The application also allows Vodafone to understand the amount of data that their customers use and as well as the trend in data usage by tracking the data usage from different applications installed on their customers' smartphones.
Such knowledge of data usage allows Vodafone to resolve connectivity issues in their network, as well as, install higher capacity links to accommodate any customer demands to support interactive applications that require higher bandwidth.
The data used by the Vodafone NetPerform mobile application is free for only Vodafone customers in Ghana, Ireland, and United Kingdom.
However, users in other countries or non Vodafone customers may be charged for any data used by the Vodafone NetPerform application.
The application is available for Android and iOS based smartphones~\cite{vodafone:googleplay, vodafone:applestore}.

\sss{Functionality} Every hour, the application establishes a TCP connection with a Vodafone server to verify whether the device has Internet connectivity.
Conducting such a test every hour allows Vodafone to understand the network stability and any variation in end-to-end latency on their network over time.
The application performs another hourly network measurement test to determine the uplink and downlink throughput against a nearby Vodafone server.
The throughput tests execute for only 10 seconds, within which the application exchanges data with a Vodafone server~\cite{vodafone:googleplay}.
The throughput is then calculated as the average of different throughput values sampled in 10 seconds.

\sss{Data Collection} The data collected by the Vodafone NetPerform mobile application is stored on Vodafone servers for only 14 months, which allows Vodafone to understand the changes in the seasonal use of the network usage by their customers.
To understand and diagnose the network problems related to phone call connectivity, the application collects cellular tower ID to which the device is connected, signal strength, device location when the network is either limited or not available, the quality of 2G/3G coverage, device speed~(if available through GPS), and time duration when the device uses cellular network, how the phone call ends~(dropped or disconnected by the user)~\cite{vodafone:terms}.

To understand and diagnose issues related to data services the application additionally captures whether the device can establish a connection with a Vodafone server, time taken to establish a connection with a Vodafone server, the MAC addresses of all available \mbox{Wi-Fi} access points along with their link bandwidth, hourly data usage of the device, data usage when the device is in standby mode, and the upload and download throughput~\cite{vodafone:terms}.

To understand the types of Internet services that users are interested in and to allocate high capacity bandwidth for services that require high bandwidth, the application captures the names of all applications installed on the device, the names of applications that the user uses everyday, the duration of application use, the amount of data is received and sent from each installed application. 

Finally, to diagnose and resolve device related network issues, the application collects the device model and company, device IMEI~(encrypted to maintain anonymity), the OS running on the device, firmware version, the OS language, battery status, memory in use, the time when the phone last rebooted~\cite{vodafone:terms}.

\sss{Resource Incentives and Protection} Users do not get any incentives for running measurement tests on their devices.
Instead, Vodafone relies on users' curiosity to understand the network performance and gathers data collected on users' devices to improve the quality of their voice and data services.
With respect to protecting device resources, the application does not allow users to configure a monthly cap on the amount of cellular and \mbox{Wi-Fi} data that the application can use to run measurement tests.
Further, since the application run throughput and latency tests every hour, the application prevents the device to turn off its radio, which drains the device's battery quickly~\cite{vodafone:googleplay}.

\sss{Privacy Protection} The application does not collect any personally identifiable information such as the device phone number, the phone numbers of incoming and outgoing phone calls, incoming and outgoing SMS messages, and the names of available \mbox{Wi-Fi} hotspots.
However, by collecting the names of application installed and when different applications are used, the Vodafone NetPerform application has a potential to discover user behavior, which might be unsuitable for some users.

\sss{Limitations} 
The Vodafone NetPerform mobile application does not allow users to discover whether their cellular ISPs are performaing traffic discrimination.
Further, the availability of the application only for users in a few countries in Europe restricts the network operators in other countries to gain insight of their network issues and performance.

}

\input{table_service.tex}

\edit{
\subsubsection{Emerging Applications}

Many other network measurement services have been developed by independent developers to assist users to measure performance of wireless and cellular networks.
Such applications include 
SpeedSpot~\cite{speedspot:homepage}, 
Sensorly~\cite{sensorly:homepage}, 
RootMetrics~\cite{rootmetrics:homepage}, 
NetworkCoverage~\cite{kaupNetSys15}, 
Internet Speed Test~\cite{internetSpeedTest:homepage}, 
Netradar~\cite{netradar:homepage}, 
Cisco Data Meter~\cite{ciscodatameter:homepage}, 
4Gmark~\cite{4gmark:homepage}, and 
nPerf~\cite{nperf:homepage}.
Because the details of how these services are implemented and how they perform measurements is not widely available, we restrict our discussion to the commonalities and differences of what data these tools collect and how.
We also illustrate the similarities and differences between these emerging measurement services  in Table~\ref{tbl:services}.

Specifically, SpeedSpot, Sensorly, RootMetrics, and NetworkCoverage are similar to the OpenSignal mobile application in their capability for users to compare the performance of their wireless and cellular networks on a map and find nearby \mbox{Wi-Fi} networks.
Internet Speed Test is similar to Ookla SpeedTest; it allows users to measure the latency and throughput to application's servers on user's network.
Similar to RadioOpt, \mbox{NetRadar} and Cisco Data Meter applications run latency and throughput experiments against servers deployed on the cloud/CDN servers and allows users to monitor traffic sent and received by applications installed on their devices.
4Gmark and nPerf are similar to each other, in that, these applications not only allow users to measure the performance of network in terms of throughput and latency, but also measures the suitability and reliability of the network for streaming and Web applications.

}

\edit{
\subsection{Network Discovery and Diagnosis}
\label{sec:services:diagnosis}

While most of the previous projects focus on measuring end-to-end performance of mobile application communications, the  following tools allow developers and researchers to learn more about the state of network infrastructure and configurations that affect transmission of application traffic.
Pertinent features include the presence of proxy servers and other middleboxes, or complex multi-level DNS resolutions.

%@@ how will you tie this into what operators care about?
%@@ the paper looks disjointed here, since operators are
%@@ not mentioned anymore. 
}

\subsubsection{NDT (Mobile Client)}
\label{sec:ndt}

The Network Diagnostic Test~(NDT) system, developed by Internet2, evaluates the performance of mobile connections to diagnose problems that limit network bandwidth~\cite{ndt:mlab, ndt:internet2}. 
NDT also detects problems associated with device misconfiguration and network infrastructure. 
NDT (Mobile) is currently hosted on Google's \mbox{M-Lab} and allows access to its backend through an Android mobile application.

\sss{Functionality} NDT measurements are performed from a mobile Web browser that issues requests to NDT servers, hosted by M-Lab.
The server-specific tests diagnose observed network problems.
%@@ what kind of network problems are diagnosed?
%@@ i thought they just identify them, they dont' have insight
%@@ into the network to diagnose them? 
%@@Utkarsh - Agree, but that is how they talk about NDT
After the measurement experiment completes, the server analyzes the results and returns them to the client device.

\sss{Data Collection}  The NDT mobile application collects traffic performance information such as upload and download speed, round trip network latency (minimum, average, and maximum), jitter, TCP receive window size (current and maximum), packet loss, TCP retransmission timer, and number of selective acknowledgements received. 
The application also detects router cable faults, incorrectly set TCP buffers in the device, duplex mismatch conditions on Ethernet links, presence of NAT, and capacity limits.

\sss{Resource Incentives and Protection} The incentive model for the NDT mobile client is based on providing network diagnostic information in exchange for users running tests on their mobile devices.
%@@ isn't this true for ALL the measurement tools?
%@@ i am quite suspicious of this section, having read it 
%@@ 20X bynow..
One issue for users who volunteer their device resources is that NDT requires permission to prevent the phone from going into power save mode, which may drain the battery quickly.
%@@ wait, does it run at all if the user doesn't provide
%@@ that permission?  don't all tools do othat? 
%@@Utkarsh - not all tools require the phone to be awake all the time. The permissions are asked at the time of app installation. So if the user does not give the permission, the app will not be installed.

\sss{Privacy Protection}  By default NDT records experimental data separately for each user, which allows users to privately diagnose their network problems.
%@@ what about the privacy sensitive information collected,
%@@ how is it protected?
%@@Utkarsh - it does not collect any privacy sensitive information.
Data isolation also prevents malicious users from learning of open ports and interfaces in others' networks.

\sss{Limitations}  
The NDT mobile client executes experiments that evaluate network traffic only between a mobile device and its closest M-Lab server and not any arbitrary server.

\subsubsection{Netalyzr}

%@@ how does the below tool help app developers develop
%@@ application?  how they deal with middleboxes?
%@@Utkarsh - not sure what you are asking for before the section starts.

The Netalyzr mobile application, developed as a collaboration of ICSI Berkeley, UC Berkeley, HIIT, and Aalto University, is a diagnostic tool that characterizes connectivity, performance anomalies, and network security issues~\cite{Kreibich:2010:NIE:1879141.1879173,aims:netalyzer}. 
The tools measures network latency and bandwidth to reveal insight into not only performance to cloud servers, but also how middleboxes in the path affect the performance of traffic.
As of March 2014, Netalyzr has run over 15000 times to diagnose 290 operators in 90 countries. 
Netalyzr is accessible via an Android mobile application available on the Google Play Store.

\sss{Functionality}  Netalyzr identifies the presence of Network Address Translations~(NATs), proxy servers along a route, IP fragmentation, size of bottleneck buffers, reachability of services, and presence of HTTP proxies. 
When the Netalyzr application starts, it contacts the Netalyzr's Web server, which issues a DNS lookup request to redirect the user's request randomly to one of the twenty Netalyzr's back-end servers hosted on the Amazon cloud. 
Each back-end server supports twelve concurrent measurement sessions.

Netalyzr detects the presence of a NAT based on a difference between a user's local and public IP addresses. 
For clients behind a NAT, Netalyzr identifies how the network renumbers addresses and ports, i.e., whether the NAT uses fixed associations of local IP addresses to different public IP addresses, or if the NAT uses load-balancing.

To detect support for IP fragmentation, Netalyzr sends a 2\,KB UDP packet (larger than 1500\,B Ethernet maximum transmission unit~(MTU)) to the server -- a response from the server indicates the network supports fragmentation. 
If there is no response Netalyzr uses binary search to find the maximum packet size it can deliver without the packet being fragmented at the IP layer.
The same test repeats from server to client to detect network support for fragmentation on the reverse path.

The sizing of bottleneck buffers affects user-perceived latency, and is measured based on the difference in latency during inactivity and during path throughput tests.
Finally, queue drain time indicates the size of the buffer.
To perform service reachability related experiments, the application attempts to connect to 25 different well known ports on a back-end server.
%@@ who is performing service reacahability experiments?
%@@Utkarsh - the client on which the application is installed.

Netalyzr infers the presence of HTTP proxies if the public IP address in the request received by the back-end server is not the same as the client's public IP address. 
To detect the presence of in-path HTTP proxy, the client first sends an HTTP request to the server, the server then returns the request headers it received in the request back to the client.
The client then compares the headers it sent and the headers the server sent to the client for any added, deleted or modified fields. 
To detect the presence of caching policies, the application relies on the HTTP 304 Not Modified response from the server.

To detect the presence of a DNS-proxy server or firewall, the application sends a DNS request to Netalyzr's back-end server.
If the client detects any change in the response (different transaction ID, or public IP address), then Netalyzr assumes an in-path DNS proxy exists. 
Netalyzr then makes invalid DNS requests to the back-end server. 
If the client receives an invalid response from the server, nothing is detected, but if the request is blocked, Netalyzer assumes a DNS-aware middlebox is blocking invalid DNS requests from leaving the network.

\sss{Data Collection}  The Netalyzr mobile app records the presence of network interfaces, gateways, NAT detection, port renumbering, path MTU, packet fragmentation, DNS resolver, extension mechanisms for~DNS~(EDNS) support, port randomization, IPv6 support, hidden proxies, in-path caches, header manipulation, image transcoding, compression, HTTP type filtering, port filtering, traffic differentiation, IP fragmentation, signal-to-noise ratio, Wi-Fi/cellular configuration, network topology through traceroute, TLS handshake, UPnP vulnerabilities on Wi-Fi APs, clock drift, and TLS default certificates~\cite{aims:netalyzer}.

\sss{Resource Incentives and Protection} Netalyzr provides network diagnostic and troubleshooting information to users.
Netalyzr requests user permission to modify system settings and to terminate other running applications in order to increase measurement accuracy.
The Netalyzr mobile application asks users for permission to execute IP traceroutes, since ICMP packet transmission on a mobile device requires access to raw sockets.

\sss{Privacy Protection} Netalyzr asks users to opt in to the data collection process before installing the application.
Therefore, if users are uncomfortable with sharing the measurement results with Netalyzr, they may not install the application.
However, when the user grants permissions to the application, Netalyzr could use GPS to get device location, read phone status and identity, and modify or delete the contents of USB storage to store or delete measurement related data on the device.

\sss{Limitations} Although Netalyzr provides a robust diagnostic set of end-to-end network measurements and helps users troubleshoot networks, unlike MITATE, or WindRider, Netalyzr does not detect traffic shaping in mobile ISPs.

%Portolan ~\cite{portolan}) @@removed because it's not a testbed and not for performance measurement, it's about (wireline) topology measurement. i think the intended audience of this paper won't care about topology measurement. and the paper is way too long already.
%@@ mwittie 5/6/15: Portolan also creates coverage maps and produces network level metrics such as throughput. However, it really first the definition of a service, so we'll move it to that section and possible remove then.

\subsubsection{PortoLan}
\label{sec:services:diagnosis:portolan}

PortoLan is a network experiment testbed based on volunteered mobile devices that executes experiments submitted to back-end servers~\cite{gregori13:sensing}.
PortoLan is designed by Enrico Gregori~\textit{et~al.} at Istituto di Informatica e Telematica, to discover Internet topology and build wide scale mobile network signal quality maps.
The Android application for PortoLan is available on Google Play and allows users to run measurement tests like ping, traceroutes, maximum throughput, and detection of traffic shaping of BitTorrent traffic~\cite{portolan:googleplay}.   
The PortoLan team intends to add capability to support active network experiments and access to mobile sensor data such as network signal strength, device location, network name, cell type, and roaming status.
PortoLan relies on user altruism to build testbed capacity and support measurement.
The PortoLan mobile application limits the device cellular bandwidth usage to 2~MB/day and postpones experimentation when battery drops below 40\%. 
Finally, the application does not collect personally identifiable information from the device and anonymously stores  measurement data on backend servers.

%% file: table_service.tex
\begin{table*}
\centering
\begin{tabular}{|>{\arraybackslash}p{3.5cm}
|>{\arraybackslash}p{1cm}
|>{\arraybackslash}p{1cm}
|>{\arraybackslash}p{1cm}
|>{\arraybackslash}p{1.4cm}
|>{\arraybackslash}p{1cm}
|>{\arraybackslash}p{2cm}
|>{\arraybackslash}p{1cm}
|>{\arraybackslash}p{1cm}
|>{\arraybackslash}p{1cm}
|}
\hline
                                                    & \rotatebox[origin=l]{90}{\textbf{SpeedSpot}}
& \rotatebox[origin=l]{90}{\textbf{Sensorly}}
& \rotatebox[origin=l]{90}{\textbf{RootMetrics}}
& \rotatebox[origin=l]{90}{\textbf{NetworkCoverage}}
& \rotatebox[origin=l]{90}{\textbf{Internet SpeedTest~}}
& \rotatebox[origin=l]{90}{\textbf{NetRadar}}
& \rotatebox[origin=l]{90}{\textbf{Cisco Data Meter}}
& \rotatebox[origin=l]{90}{\textbf{4GMark}}
& \rotatebox[origin=l]{90}{\textbf{nPerf}}
\\ \hline
\multicolumn{10}{|l|}{\textbf{Support measurement tests}}                                                                                                                                                                                                                                                                                                                                                                                                                          \\ \hline
Uplink throughput                                                     & \cm                                                        & \cm                      & \cm                   &                                                                       & \cm                         & \cm                                                                  & \cm                                                     & \cm                      & \cm                      \\ \hline
Downlink throughput                                                   & \cm                                                        & \cm                      & \cm                   & \cm                                                                   & \cm                         & \cm                                                                  & \cm                                                     & \cm                      & \cm                      \\ \hline
Latency                                                               & \cm                                                        & \cm                      &                       & \cm                                                                   & \cm                         & \cm                                                                  & \cm                                                     & \cm                      & \cm                      \\ \hline
Signal coverage maps                                                  &                                                            & \cm                      & \cm                   & \cm                                                                   &                             & \cm                                                                  &                                                         &                          &                          \\ \hline
\multicolumn{10}{|l|}{\textbf{Uplink throughput tests}}                                                                                                                                                                                                                                                                                                                                                                                                                            \\ \hline
Total uplink transfer                                                 & 6\,MB                                                      & 10\,MB                   & 7\,MB                 &                                                                       & 8\,MB                       & Random                                                               & 900\,KB                                                 & 50\,MB                   & 20\,MB                   \\ \hline
Probe method                                                          & P                                                  & P                & P             &                                                                       & P                   & TCP                                                                  & P                                               & P                & P                \\ \hline
No. of probes                                                         & 1                                                          & 2                        & 1                     &                                                                       & 12                          & Random                                                               & 1                                                       & 1                        & 20                       \\ \hline
Duration                                                              & 6\,s                                                  & U & 8\,s             &                                                                       & 15\,s                & 10 \,s                                                           & U                                & 10\,s               & U \\ \hline
\multicolumn{10}{|l|}{\textbf{Downlink throughput tests}}                                                                                                                                                                                                                                                                                                                                                                                                                          \\ \hline
Total downlink transfer                                               & 20\,MB                                                     & 400\,MB                  & 2400\,MB              & 10\,MB                                                                & 400\,MB                     & 1\,MB                                                                & 2\,MB                                                   & 250\,MB                  & 10\,GB                   \\ \hline
Probe method                                                          & G                                                   & G                 & G              & G                                                              & G                    & TCP                                                                  & G                                                & G                 & G                 \\ \hline
No. of Probes                                                         & 2                                                          & 4                        & 4                     & 1                                                                     & 4                           & 16                                                                   & 2                                                       & 1                        & 10                       \\ \hline
Duration                                                              & U                                   & 10\,s               & 25\,s            & U                                              & 15\,s                  & 10\,s                                                           & U                                & 10\,s               & 10\,s               \\ \hline
\multicolumn{10}{|l|}{\textbf{Latency tests}}                                                                                                                                                                                                                                                                                                                                                                                                                                      \\ \hline
Probe Method                & G                                                   & G                 &                       & T                                                              & P                   & G                                                             & T                                                & G                 & G                \\ \hline
No. of probes                                                         & 10                                                         & 6                        &                       & Random                                                                & Random                      & 2                                                                    & 30                                                      & 3                        & 10                       \\ \hline
\multicolumn{10}{|l|}{\textbf{Measure Application performance}}                                                                                                                                                                                                                                                                                                                                                                                                                    \\ \hline
Webpage load time
& \multicolumn{1}{l|}{}                                      & \multicolumn{1}{l|}{}    & \multicolumn{1}{l|}{} & \multicolumn{1}{l|}{}                                                 & \multicolumn{1}{l|}{}       & \multicolumn{1}{l|}{}                                                & \multicolumn{1}{l|}{}                                   & \multicolumn{1}{l|}{\cm} & \multicolumn{1}{l|}{\cm} \\ \hline
\begin{tabular}[c]{@{}l@{}}Throughput\\ of video streams\end{tabular} & \multicolumn{1}{l|}{}                                      & \multicolumn{1}{l|}{}    & \multicolumn{1}{l|}{} & \multicolumn{1}{l|}{}                                                 & \multicolumn{1}{l|}{}       & \multicolumn{1}{l|}{}                                                & \multicolumn{1}{l|}{}                                   & \multicolumn{1}{l|}{\cm} & \multicolumn{1}{l|}{\cm} \\ \hline
\multicolumn{10}{|l|}{\textbf{Application deployment}}                                                                                                                                                                                                                                                                                                                                                                                                                             \\ \hline
Experiments run against servers hosted by                                                   &  MaxCDN, EdgeCast CDNs & OVH, Digital Ocean CDNs     & Amazon AWS            & Think Broadband, Emanics Lab & v-speed                  & Amazon AWS, CacheFly CDN                                                & Akamai CDN                                                  & 4Gmark                   & nperf                    \\ 
\hline
Supported mobile OS platform                                                 & Android, iOS                                               & Android, iOS             & Android, iOS          & Android                                                               & Android                     & Android, iOS, Windows, MeeGo, Symbian, NokiaX, Jolla, and Blackberry & Android, iOS                                            & Android, iOS             & Android, iOS             \\ 
\hline
\multicolumn{10}{|l|}{\textbf{Misc.}}                                                                                                                                                                                                                                                                                                                                                                                                                                              \\ \hline
Developed by                                                          & Speed Spot                                                  & Sensorly                 & Root Metrics           & Technische Universität Darmstadt                                      & V-Speed                     & Aalto University                                                     & Cisco Systems & Trois Petits Points      & nPerf                    \\ \hline
Released in                                                           & May 2013                                                   & Aug. 2012                & Aug. 2011             & Mar. 2015                                                             & Oct. 2014                   & Feb. 2013                                                            & May 2013                                                & May 2013                 & Oct. 2014                \\ \hline
Number of app installs                                                & $>$ 100\,K                                                   & $>$ 50\,K                 & $>$ 100\,K              & $>$ 100                                                               & $>$ 1\,M                      & $>$ 10\,K                                                             & $>$ 50K                                                 & $>$ 500\,K                 & $>$ 50\,K                  \\ \hline
\end{tabular}\\~\\
{\raggedright
~\\
\textbf{Legend:}\\
U -- Until transfer completes.\\
G -- HTTP GET.\\
P -- HTTP POST.\\
T -- TCP Connection Setup.\\
}
\caption{Experimentation flexibility matrix of emerging end-to-end measurement services.}
\label{tbl:services}
\end{table*}

%% file: conclusions.tex
\section{Conclusions}
\label{sec:conclusions}

This survey provides a comprehensive overview of the existing and emerging end-to-end mobile network measurement testbeds, tools, and services.
In spite of the relative maturity of existing platforms, several functionality gaps remain with respect to the needs of developers, researchers, network operators, and regulators in assessing mobile network performance.
First, existing tools do not adequately support detection of traffic shaping.
As depicted in Table~\ref{tbl:exp-tools}, testbeds such as MITATE, Seattle, PhoneLab, PortoLan, and WindRider can detect the presence of traffic shaping mechanisms in mobile ISPs, whereas, other testbeds do not.
Second, device churn inherent in platforms based on ad-hoc user participation means that existing tools are not well-suited for long-term network performance monitoring.
In fact, the popularity of tools such as PhoneLab and FCC has declined over time.
Third, several testbeds enable developers to prototype the performance of their applications ahead of deployment.
However there is significant disparity in how testbeds provide that functionality in terms of execution models and APIs.
Finally, exchange of P2P traffic, network diagnostics, ICMP traceroutes, device selection criteria, and NAT traversal are not only selectively supported by different platforms.
\edit{
One significant axis of comparison between network measurement platforms not discussed in this survey is their accuracy.
The variety of measurement methods used to obtain even the relatively standard network metrics, such as throughput, makes it difficult to compare the relative accuracy of the different platforms.}

% \todo{mwittie 3/14/15: one of the reviewers suggested comparing the accuracy of the different tools. Maybe pitch it as future work since clearly that would be a significant engineering effort and results likely controversial.
% Another part of future work could be the evaluation of QoE measurement tools, though I don't think there is enough work in that space for a survey, or even a section in this one.
% Maybe mention WProf in that space (analysis of critical path), GUM in terms of objective metrics, Pangolin?, maybe a few other papers?}

Based on the surveyed work, we believe the mobile network measurement community needs a more concerted effort among developers, researchers, network operators, and regulators to produce network measurement tools that meet the needs of all four communities.
A more concerted effort would lead to greater adoption of (perhaps fewer) tools, as well as large-scale and long-term network monitoring. 
At the same time, funding agencies should support development of new measurement approaches and capabilities, especially when such improvements are aimed at enhancement of existing testbeds.

%% file: acknowledgements.tex
\ifCLASSOPTIONcompsoc
  % The Computer Society usually uses the plural form
  \section*{Acknowledgments}
\else
  % regular IEEE prefers the singular form
  \section*{Acknowledgment}
\fi

The authors would like to thank
Justin Cappos, Geoffrey Challen, David Choffness, Nick Feamster, Walter Johnston, Valerio Luconi, Jim Martin, Konstantina Papagiannaki, John Rula, Mario Sanchez, Qing Yang, and Hongyi Yao
for suggested improvements to an early version of this manuscript.